**Van der Waals heterostructures for high-performance device applications: challenges and opportunities**

Shi-Jun Liang#, Bin Cheng#, Xinyi Cui* and Feng Miao*

Dr. S. J. Liang, Dr. B. Cheng, Prof. F. Miao
[1]National Laboratory of Solid State Microstructures, School of Physics, Collaborative Innovation Center of Advanced Microstructures, Nanjing University, Nanjing 210093, China
Corresponding author: miao@nju.edu.cn
Prof. X. Y. Cui
[2]State Key Laboratory of Pollution Control and Resource Reuse, School of the Environment, Nanjing University, Nanjing 210046, China
Corresponding author: lizzycui@nju.edu.cn

#The authors S. J. Liang and B. Cheng contributed equally to this work

Abstract: Discovery of two-dimensional materials with unique electronic, superior optoelectronic or intrinsic magnetic order have triggered worldwide interests among the fields of material science, condensed matter physics and device physics. Vertically stacking of two-dimensional materials with distinct electronic and optical as well as magnetic properties enables to create a large variety of van der Waals heterostructures. The diverse properties of the vertical heterostructures open up unprecedented opportunities for various kinds of device applications, e.g. vertical field effect transistors, ultrasensitive infrared photodetectors, spin-filtering devices and so on, which are inaccessible in the conventional material heterostructures. Here, we review the current status of vertical heterostructures device applications in vertical transistors, infrared photodetectors and spintronic memory/transistors. The relevant challenges for achieving high-performance devices are presented. We also provide outlook on future development of vertical heterostructure devices with integrated electronic and optoelectronic as well as spintronic functinalities.

# 1. Introduction

The discovery of graphene (Gr) brings physicists and semiconductor engineers to a brand-new world of two-dimensional (2D) materials[1, 2]. As a monolayer of graphite, graphene has a unique electronic band structure in which low-energy quasiparticle can be described as a massless Dirac fermion[3], and exhibits novel electronic transport behaviors, such as half-integer quantum Hall effect[4], Klein tunneling[5], Dirac fluid[6, 7], etc. Moreover, graphene layer can be stacked layer-by-layer due to weak van der Waals (vdW) interlayer interaction. Electronic band structures of hybrid artificial materials strongly depend on the number of stacking layers and stacking manner, *e.g.* stacking order (A-B-C or A-B-A stacking)[8-10] and twist-angle[11, 12] . Beyond graphene, the family of 2D materials has been expanding, including semiconductors (*e.g.* transition metal dichalcogenides (TMDs)[13-20], black phosphorous (BP)[21]), insulators (*e.g.* hexagonal boron nitride (h-BN)[22]), ferrimagnets[23, 24] (*e.g.* $Gr_2Ge_2Te_6$ [25, 26], $Fe_3GeTe_2$ [27, 28] and $CrI_3$ [29]), and various topological materials (*e.g.* monolayer $WTe_2$ as topological insulator [30, 31]), and so on. With dangling-bond-free surface and weak vdW interlayer interactions among layered structures, random combinations of 2D materials can be chosen to stack in arbitrary sequence to create numerous types of vdW heterostructures[32, 33]. Not limited to the properties of individual 2D materials, the vdW heterostructures can combine superior properties of individual components to achieve diverse functions [30, 32, 34-39].

The vdW heterostructures [23, 40-49] have attracted widespread attention from the fields of materials science, condensed matter physics and semiconductor devices. They have been widely used to probe many-body physical behaviors [50-52], such as wigner crystallinzation, excitonic superfluidity, superconductivity and itinerant magnetism, which originate from strong electrons correlation between graphene layers electrically isolated by h-BN. Many degrees of freedom are available for tuning electronic properties of vdW heterostructures, such as material selections in stacking, twisted angle of component layers[50, 53] etc. In the Gr/h-BN/Gr vertical vdW heterostructure, suitable alignment of crystallographic orientation in two graphene layers allows for realization of resonant tunneling phenomenon [54, 55]. More interesting physics phenomena would emerge due to enhanced electron orbitals coupling in adjacent 2D materials when two individual 2D materials are stacked together. For example, moire patterns can be created in the Gr/h-BN vdW heterostructure[5]. Gate-tunable Mott-like insulating and superconducting as well as ferromagnetism behaviors can be observed in the twisted bi-layer electronic systems [12, 56, 57]. In addition, topological insulators[58] and shear solitons [59] have been also theoretically predicted in twisted TMD/TMD vertical vdW heterostructures. Other types of vdW heterostructure devices have been demonstrated by stacking atomically flat 2D crystals on top of other materials exhibiting ferroelectricity, magnetism and strong spin-orbit coupling [60-69], in which the proximity effect is used to alter electronic structures of 2D crystals.

The creation of a large variety of vdW heterostructures with free of lattice mismatching opens up an unprecedented opportunity for development of novel electronic, optoelectronic and spintronic devices with desirable functions and performance [69-75], as shown in Fig. 1. The ultrashort charge transport path determined by 2D materials thickness [76] in the vertical heterostructures enables to construct ultrafast switching speed and high-responsivity photodetection. Utilizing layer-by-layer antiferromagnetic ordering in atomically-thin crystal allows for realizing high-density magnetic information storage devices [73]. The functions of these diverse device applications can be enhanced by the properties of gate-tunable band offset and magnetic ordering as well as carrier density. The vdW interaction is not exclusive to 2D/2D structure, it can be extended to any passivated or dangling-bond-free surfaces. Much effort has been devoted to mixed-dimensional vdW heterostructure, including 0D/2D and 1D/2D as well as 3D/2D heterostructures, which show promise in logical devices, photodetectors and photovoltaics, light-emitting devices, energy related devices and so on. The recent progress and challenges as well as opportunities on mixed-dimensional vdW heterostructures have been reviewed in recent work[77-79].

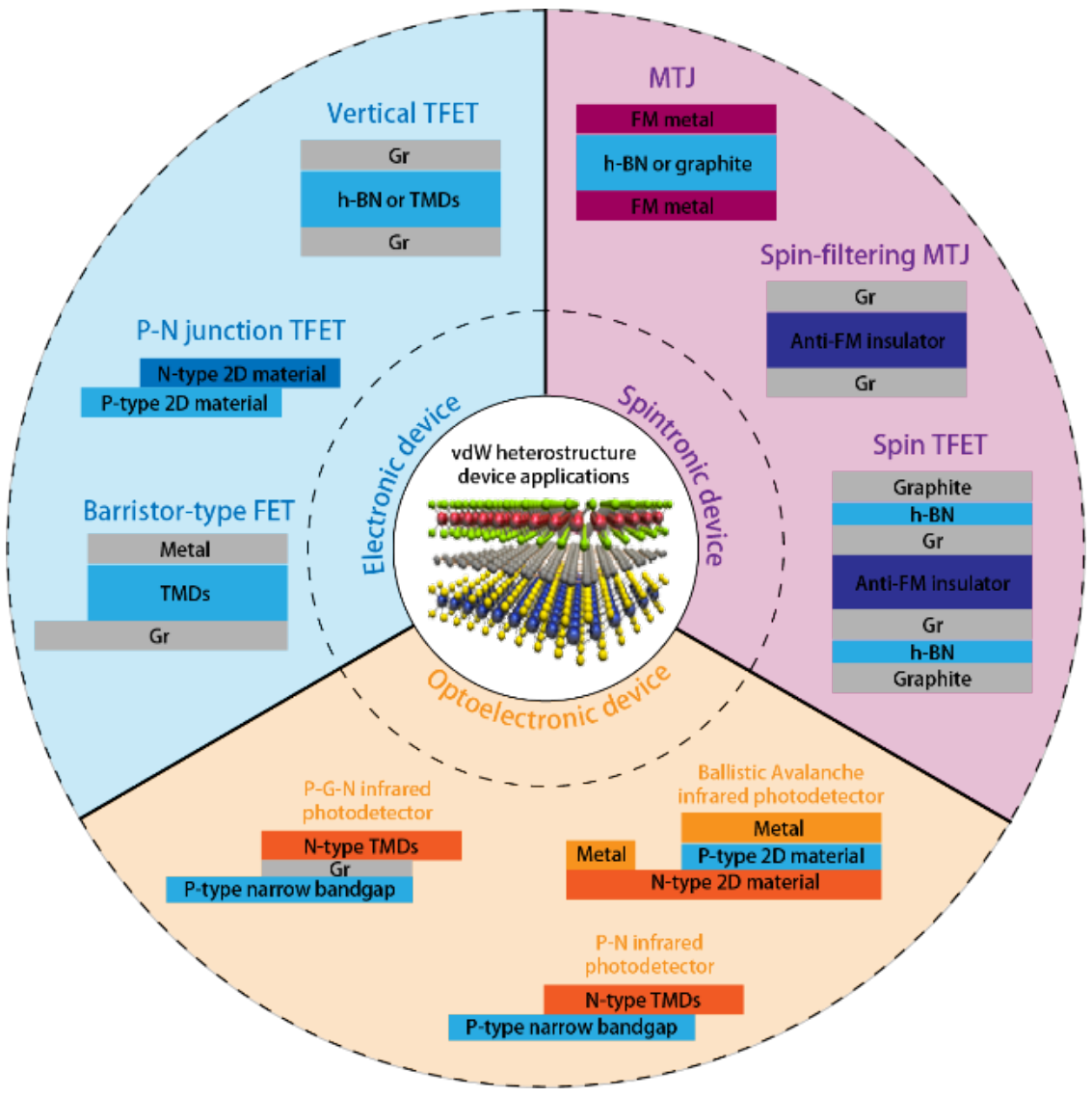

**Fig. 1** Diverse device applications based on vdW heterostructures

Vertical heterostructures have been proposed as a platform to construct vertical transistor devices with distinct operating principles from lateral field effect transistors, such as tunneling field effect transistors [80, 81], band-to-band tunneling transistors [82-85], barristors and hot electron transistors[86-88]. The early study of electronic devices focused on use of two-dimensional materials as channels, in the hope of solving the issues faced by silicon transistors. However, the performance of 2D semiconducting materials based lateral transistors usually suffers from large contact resistance that arises from high Schottky barrier at the interface of metal electrodes and 2D semiconducting materials [89-91]. Compared to the conventional silicon transistors suffering from high power consumption, the tunneling field effect transistors based on vertical heterostructures have potential in reducing supply voltage to achieve low-power consumption operation. The first reported tunneling field effect transistor is based on Gr/h-BN/Gr vdW heterostructure[92]. Its low tunneling current density ($pA/\mu m^2$) and ON/OFF ratio (around 50) can be increased to $\mu A/\mu m^2$ and $10^6$, respectively, by replacing h-BN with 2D semiconductors of smaller bandgap, *e.g.* $WS_2$ [80]. With the similar device structure, twisting the crystallographic orientation between top and bottom bi-layer graphene electrodes leads to resonant tunneling transistors [54, 83]. 2D materials can retain sharp band edges even at their atomically thin limits. This unique property can be used to create vertical p-n junction devices [93], such as $MoS_2/WSe_2$ [94], $WSe_2/SnSe_2$ [84], $BP/SnSe_2$ [82, 85] and so on, which are based on the mechanism of band-to-band tunneling. Different from the conventional planar p-n junction, the interlayer band alignment in the vertical band-to-band tunneling transistors can be tuned efficiently. By utilizing the good modulation of carriers' transport across the vertical p-n junction, a sub-threshold swing smaller than thermionic limit has been achieved over a wide drain current range in the $MoS_2$/Ge tunneling field effect transistor [95]. The tunneling transistors based on vertical heterostructure indeed show great promise in reduction of threshold swing and supply voltage, but have also some limits in the tunneling current density and ON/OFF ratio.

In addition to the electronic device applications, excellent optoelectronic properties of 2D materials also make vertical heterostructures promising in the achievement of photodetection with ultrahigh sensitivity, ultrafast response speed and polarization sensitive, especially for infrared photodetection[96-102]. Photodetection refers to a physical process that a light signal is converted into an electrical signal through a device. So far, infrared photodetection has been widely deployed in the night vision, quality inspection, optical communications and so on. These high-performance infrared photodetectors [103] are based on epitaxial-grown materials, such as HgCdTe, InGaAs and so on. However, the fabrication processes of these infrared photodetectors are very complex. Besides, requirement for cryogenic temperature operation and CMOS incompatibility issues may limit their further development and applications. Compared to these traditional materials, 2D vdW materials with many unique advantages can offer a promising and competitive platform for realizing room-temperature infrared photodetection with ultrahigh sensitivity, ultrafast response speed and polarization sensitive. The materials matrix for 2D materials is very huge and has been continuously expanding, which includes many narrow-bandgap semiconductors with anisotropic electronic and optical properties. The bandgap of these materials is tunable by varying thickness. These unique properties can be used to realize broadband infrared response with polarization. Enhanced light-matter interactions give rise to strong optical absorption in the atomically thin vdW heterostructures, which arises from the presence of van Hove singularities in the electronic density of states of 2D materials[104]. Different doped 2D materials (*i.e.* n-type and p-type) can be found in this large material database. Fabricated vertical p-n vdW junctions[93] not only have ultrashort transport channel for photogenerated carriers to achieve ultrafast response speed, but also show very strong built-in electrical field for accelerating exciton dissociation and suppressing dark current to realize ultrahigh sensitivity. Together with these unique and excellent properties, dangling-bond-free surface of vdW heterostructures can be transferred to arbitrary substrates to achieve on-chip integrated electronic and optoelectronic as well as magnetic devices for practical applications[105].

With recent discovery of 2D intrinsic magnetic vdW crystals [27-29, 106], additional functionality can be added to vdW heterostructure devices to provide more exciting opportunities for disruptive device applications [73, 107-109]. Recently reported pristine 2D materials include ferromagnetic semiconductors/metals, antiferromagnetic insulators and so on[23, 24]. The magnetism of these 2D magnets can be tuned either through applying electrical field or varying carrier concentration[26, 110-112]. Such properties may fuel possibilities for devising novel magnetoelectric devices such as memory devices, logic devices, etc. Unlike the conventional magnetic tunneling junction currently used in spintronic industry as a building block, vdW MTJ devices based on 2D magnetic insulator gives rise to a huge magnetoresistance exceeding $10^4$ at low temperature [73, 113, 114], which shows great promise of artificial vdW magnets in the information storage. Similar to other 2D nonmagnetic crystals, 2D vdW magnets are able to realize seamless integration with other layered materials with interesting electronic and optical properties, providing tremendous freedom for constructing heterostructures to study attractive physics properties and achieving more device applications beyond conventional electronic and optoelectronics. In the bilayer heterostructure consisting of nonmagnetic and magnetic materials, proximity effect not only can be used as a tool to engineer the electronic structure of 2D nonmagnetic material [69, 115-119], but also allows for exploration of novel spintronics applications. The control of such vdW heterostructure devices may become more diverse as the magnetism, electronic and optical properties of components in vdW heterostructures can be dynamically tuned by the electrical, strain and light and thermal approaches.

There are still some challenges to overcome, although vdW heterostructure applications in electronic, infrared photodetectors and spintronics show many advantages over the conventional devices. For examples, extremely small subthreshold swing (3.9 mV/dec) can be reached in the tunneling field effect transistor based on vertical p-n junctions[95], the available on-current density is far below the requirement for practical applications. The highly-sensitive infrared photodetectors based on narrow-bandgap semiconductors, *e.g.* BP, B-AsP, have been achieved. Nevertheless, these materials are air-sensitive and very unstable in the ambient conditions. Additionally, the achievement of high-sensitivity in the infrared photodetection is at the cost of slow response speed [120]. For the vdW heterostructures based spintronics applications, electrically tuned interlayer magnetic order promises the device concepts such as spin-filter magnetic tunneling junction [73] and spin field effect transistor [108], however use of a large magnetic field and requirement for low temperature limit their practical applications in non-volatile memory devices and logic devices.

In this review, we summarize the device applications based on vdW heterostructure as shown in Fig. 1. We start with an overview of the research progress along the application of vertical heterostructure in the electronic devices and their working principles. In the second section, we are going to review the relevant works about infrared photodetection based on diverse vdW heterostructures and discuss the figures of merit and applications. In the third section, the recent work on vdW-based 2D spintronic devices will be summarized. Finally, we will discuss the challenges faced by vertical transistors, infrared photodetectors and spintronic devices based on vdW heterostructures and provide the outlook on future development.

## 2. Vertical vdW heterostructures for field effect transistor

Field effect Transistors (FETs) based on two dimensional materials give new opportunities for overcoming the shortcomings of the traditional FETs. Especially, it is expected that the short channel effect can be resolved by the nature of 2D materials, which is exemplified by the demonstration of $MoS_2$ FET with 1-nm gate length [121]. Although large current [122] in graphene device was achieved, but ON/OFF ratio is less than 10 due to its semimetal nature[123-125]. The issue on low ON/OFF ratio can be solved with the discovery of semiconducting TMDs. Bulk semiconducting TMDs show direct bandgap around 1.2-1.4 eV and transform into direct bandgap in their monolayer form with 1.8-2 eV. FETs based on single layer $MoS_2$ exhibit an ON/OFF ratio up to $10^8$ [126, 127]. For monolayer semiconducting TMDs based FETs, their ultimate performance and potential applications are limited by the large contact resistance at the interface of metal electrodes and 2D semiconductors[89-91]. The ON-current density in monolayer 2D materials based planar FETs is far below that of silicon devices. By using strageties like phase engineering, transferred electrodes and inserting tunneing barrier and so on, the performance has been improved, but it remains difficuilt to meet the requirement for practical applications (above 0.8 ~1.6 mA/μm[128]). Recent efforts have delivered record-high ON-current densities in n-type $MoS_2$ (0.83 mA/ μm[128]) and p-type BP (1.2 mA/μm[129]) by shrinking channel length down to sub-100 nm and improving contact resistance[130], which is comparable to that of silicon delvices. As an alternative path to high-performance planar FETs, vertical FETs based on vertical vdW heterostructures of 2D materials have attracted lot of attentions[80, 93, 131-135]. Different types of vertical vdW FETs have demonstrated by using distinct electronic properties of 2D materials, such as gate-tunable Schottky barrier at the interface of Gr/TMD, excellent dielectric of h-BN and gate-tunable band alignment of atomically thin p-n junction. Gate-tunable interface barrier enables transition between tunneling and thermionic processes in vertical FETs. The vertical FETs improve the performance over the planar devices in OFF current and subthreshold swing as well as channel length [136]. Below, we will give several examples to show their advantages and some unavoidable shortcomings.

**2.1 Gr/h-BN/Gr vertical tunneling FETs**

Recently, Dirac-source planar FET has shown a room-temperature subthreshold swing (~40 mV $dec^{-1}$) much lower than the theoretical limit (60 mV $dec^{-1}$) imposed by thermal Boltzmann excitation, due to quickly reduced "thermal tail" of the electron density [137, 138]. The achievement of such low subthreshold swing is due to the use of graphene's linear band structure, in which the density of states super-exponentially drops as the Fermi energy increases, leaving less electrons above the tunneling barrier. In the graphene based tunneling vertical junction, such an advantage can be inherited. Moreover, strong quantum capacitance effect in monolayer graphene allows for Fermi level tuning with gate over a large range[86]. As a result, it is a good choice to use graphene as electrode for the vertical tunneling FETs.

The first tunneling transistor based on vertical vdW heterostructures of 2D materials was realized in Gr/h-BN/Gr structures, in which top and bottom graphene layers serve as electrodes, while the h-BN acts as a tunneling barrier[92] (Fig. 2a). The h-BN with 4~7 layers (1.6 ~ 2.8 nm) has been selected to optimize the field effect. The screening effect on monolayer graphene is very weak due to the low carrier concentration in the bottom graphene, so the Fermi level of both top and bottom graphene electrodes can be effectively tuned. However, the large effective tunneling barrier leads to small ON/OFF ratio ~ 6 (50) for electrons (holes) at room temperature (Fig. 2b), which can be seen from band alignment shown in Fig. 2c, d and e. In contrast to graphene planar transistor, the ON/OFF ratio in vertical tunneling transistor is almost temperature-independent, because the barrier height of the h-BN layer is much larger than the thermal activation energy of room temperature[139]. The h-BN barrier layer can be replaced by $MoS_2$ layer, in which the effective barrier height is much lower than h-BN and comparable with the reachable Fermi level in monolayer graphene by gate tuning, so that the current in the OFF state in the transistor is tunneling type while the current is ohmic in the ON state. As a result, the current ON/OFF ratio can be as high as ~10,000, which is large enough for achieving logic circuits. Compared to the graphene planar transistor, another advantage of graphene tunneling transistor is the much shorter channel length, which is promising in vertical integration of circuits.

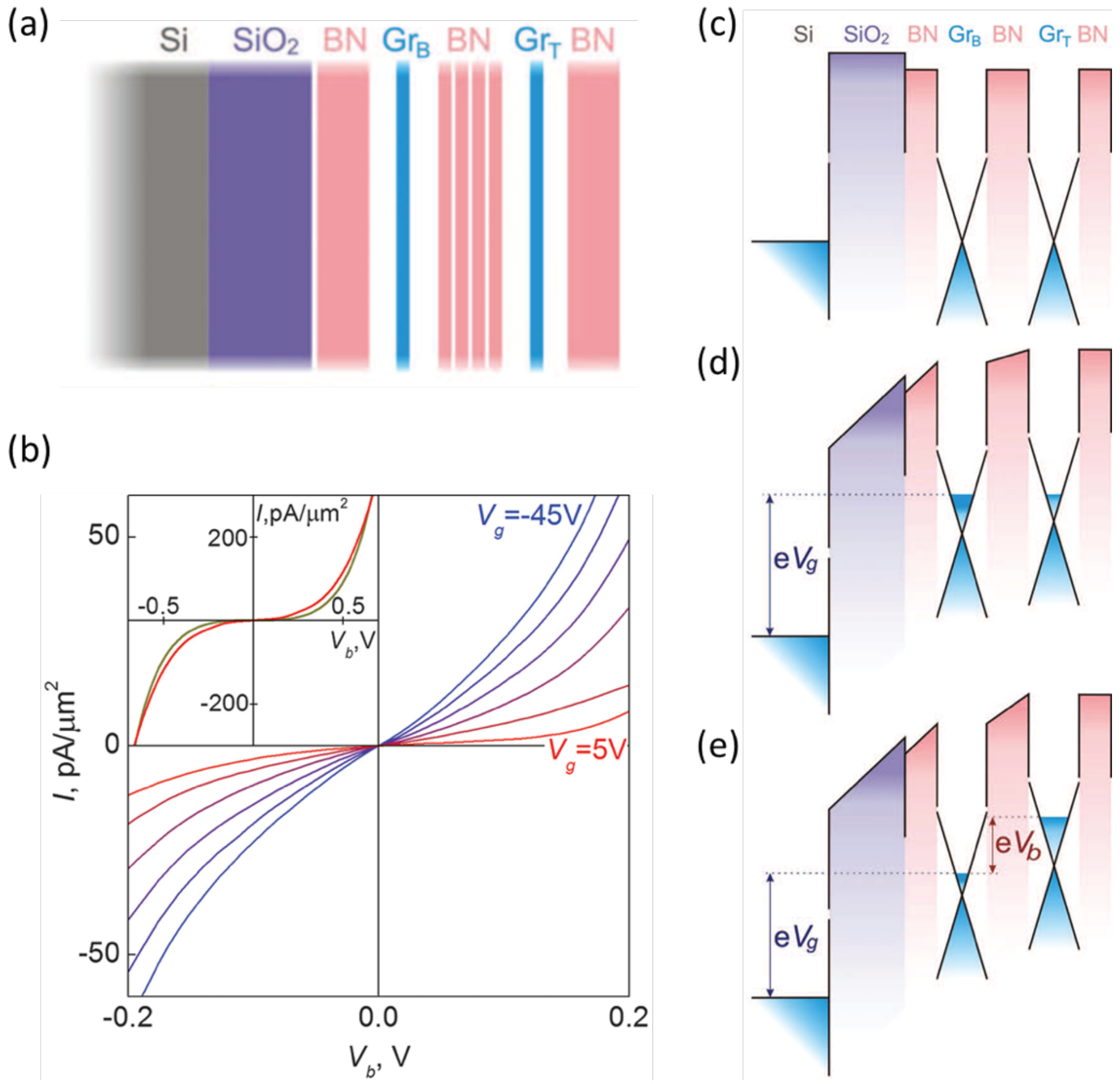

**Fig. 2** Tunneling FET based on Gr/hBN/Gr vetical heterostructures. (a) The schematic image of the device. Copyright (b) I-V characteristics for different gate voltages and the inset shows a comparison between experiment (red line) and theory (dark line). The band alignment with (c) zero bias and gate voltage, (d) zero bias and non-zero gate voltage, and (e) non-zero bias and gate voltage. Reproduced with permission[92]. Copyright 2012, The American Association for the Advancement of Science (AAAS).

Use of thick h-BN tunnel layer (4~7 layers) may kill the electron tunneling and leads to small ON current density (as low as tens of pA/μm$^2$), although it is possible in theory to achieve high ON/OFF ratio (exceeding $10^4$) with such thickness[92]. It is critical to increase the ON current density while keep the high ON/OFF ratio simultaneously for improving the performance of graphene-based vertical tunneling FET. When some interlayer scatterings between the top and bottom graphene layers are considered, the tunneling probability can be enhanced. It is worth to be noted that the tunneling current can reach tens of nA/μm$^2$ which is about 1000 times large than the tunneling current without any resonant process. According to the famous Bardeen transfer Hamiltonian approach, the tunneling current caused by interlayer scattering is proportional to the difference of Fermi energies, the density of states, and the scattering potential. When scattering potential V($\boldsymbol{q}$) has bell-shaped dependence on wave vector $\boldsymbol{q}$, i.e. V($\boldsymbol{q}$) has a maximum value at $\boldsymbol{q} = \boldsymbol{q_c}$, the I-V curve will have a resonant peak at bias $V_b \approx \hbar v_F q_c / e$. [134]. As shown in Fig. 3, the low density of states dominates the low bias region in I-

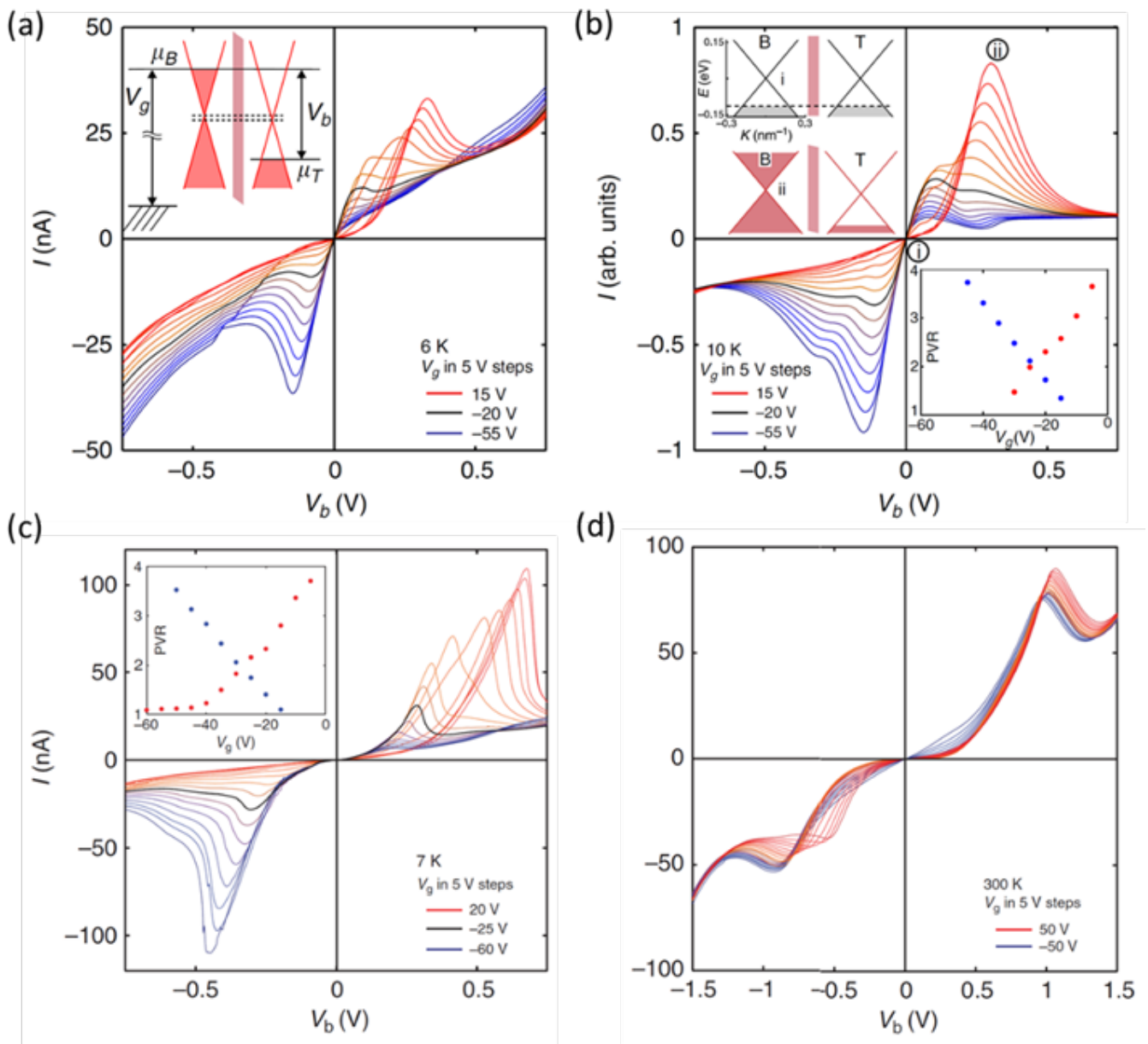

**Fig. 3** Resonant tunneling in graphene/hBN/graphene tunneling FET. (a) measured and (b) theoretical I-V curve for different gates. The band alignment for different regions is illustrated in the insets. (c) (d) Resonant tunneling for different gates at 7K and room temperature, respectively. Reproduced with permission[134]. Copyright 2013, Springer Nature.

V curve, while the mismatch of wave vector in the scattering process gives rise to a current drop in the high bias region, leading to negative differential conductance (NDC). The tunneling behavior is gate-tunable since the Fermi level of bottom layer and top layer graphene can be simultaneously changed. As a result, the electronic states at the Fermi level contribute to the tunneling current and cause asymmetry in the I-V curve, according to Fermi's Golden rule. As the temperature increases to room temperature, the improved ON current of resonant tunneling and NDC phenomenon are still retained, although the peak-to-valley ratio drops from 4 to a relatively small value, indicating possible applications in high-speed electronics.

The resonant scattering processes indeed help to improve performance in TFET based on graphene, nevertheless, the ON current density remains too small. The reason is that a large momentum transfer is required for interlayer scattering between two mismatched wave vectors in top and bottom graphene electrodes. When the crystallographic orientations of the top and bottom graphene layers are only slightly twisted, it is possible that the conservation of momentum (or wave vector ***k***) takes places without any scattering processes in the carrier transport process [55], and that the bias voltage for the resonant peak of I-V curve is only proportional to the twisted angle (Fig. 4 a-e). In this case, the scattering sources broaden the resonant peak instead of serving as "bridge" for the tunneling, and the ON current density can reach several hundred of nA/μm$^2$ (Fig. 4 f-g). The momentum conservation can be further confirmed as the in-plane magnetic field is applied, where an additional term related to the tunneling distance and the twisted angle emerges (Fig. 4 h-i). On the other hand, if the monolayer graphene electrodes are replaced by bilayer graphene, the resonant peak and NDC also show up [140]. Moreover, when the resonant tunneling occurs in that device, the energy conservation is strictly retained, while the requirement of momentum conservation is largely-loosened compared to the case of monolayer graphene. The reason for such loosened momentum conservation is that the bilayer graphene has quasi-parabolic energy-momentum dispersion with a nearly-flat band edge. At the small twisted angle, conservation of energy is satisfied within a large range of wave vector ***k***, which is proportional to the energy difference of the neutral charge points in the two separate bilayer graphene, determining the broadening of the resonant peaks.

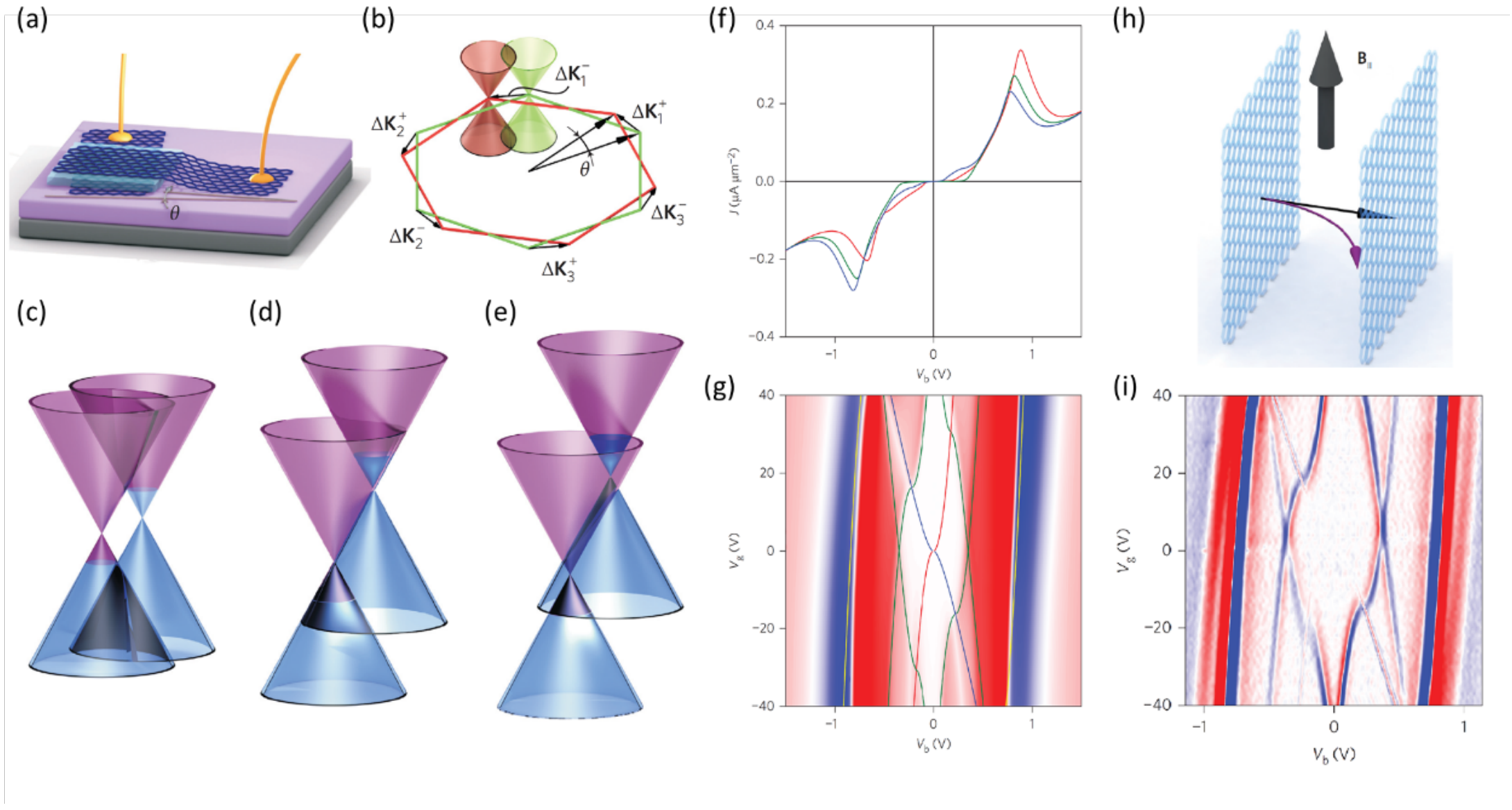

**Fig. 4** Resonant tunneling in small twist-angle Gr/h-BN/Gr tunneling FET. (a) Schematic image of device and (b) band structure. (c) (d) (e) Band alignment for different bias voltage before, at and beyond resonant tunneling regions. (f) I-V curve at different gate voltage. (g) different conductance vs. bias and gate voltages. (h) Schematic image of in-plane magnetic field. (i) Effect of resonant tunneling by the in-plane magnetic field. Reproduced with permission [55]. Copyright 2014, Springer Nature.

### 2.2 Gr/TMD/Gr vertical tunneling FETs

In the vertical tunneling TFETs, large tunneling current and high ON/OFF ratio are like two sides of the same coin and cannot be satisfied simultaneously. New operating mechanism is required to increase the ON current by several orders in the tunneling FET. For Gr/h-BN/Gr vertical TFETs, the energy different between Dirac point in the graphene and upper edge of valence band of h-BN is much larger than tunable-Fermi level range by gating in graphene. As a result, the tunneling mechanism dominates over all gate tuning range. To increase the ON current, it is desirable to have energy difference comparable to or smaller than tuning range of Fermi level by gating at ON state. Compared to the wide bandgap of h-BN, TMDs show much smaller band gap and can be used to replace h-BN as potential barrier in the vertical TFETs. The indirect bandgap of bulk $MoS_2$ (or $WS_2$) is ~1.23 eV (or ~1.35 eV) and increases to ~1.8 eV (or ~1.9 eV) in its monolayer form with direct gap. The electron affinity of bulk $MoS_2$ (or $WS_2$) is 4.5 eV (or 4.1 eV) and reduces down to 4.3 eV (or 3.9 eV) in monolayer materials[141]. Using $MoS_2$ indeed increases the ON current density but it is still small for practical applications (~0.1 μA/μm$^2$) [92]. Different from the position of valance band edge in $MoS_2$, the upper edge of valence band of $WS_2$ is almost in alignment with graphene Dirac point without any gating [80]. Because the linear energy-momentum dispersion, the Fermi level can be tuned over a wider range than that in $WS_2$. As a result, when the gate is turned on to lower the Fermi level of the top and bottom graphene to the valence band, the tunneling barrier height of the vertical heterostructure is increased, leading to a reduced tunneling current. In contrast, when the Fermi level in the graphene is tuned to align with the conduction band of $WS_2$, the barrier height is drastically lowered or even overpassed, resulting in over-barrier thermionic current, which is determined by the thermionic activation to overcome the chemical potential drop between the bottom and top graphenes, rather than tunneling current (Fig. 5 a-d). Therefore, the ON current ($2\times10^{-6}$ A/μm$^2$) and ON/OFF ratio (exceeding $10^6$ at room temperature) in Gr/$WS_2$/Gr vertical device are much larger than that in Gr/$MoS_2$/Gr and Gr/h-BN/Gr devices (Fig. 5e-f). However, the lowered tunneling barrier in the vertical tunneling transistor is not large enough to block the thermal excitation across the barrier at the OFF state, leading to strong temperature dependence of performance of the tunneling transistor (Fig. 5g).

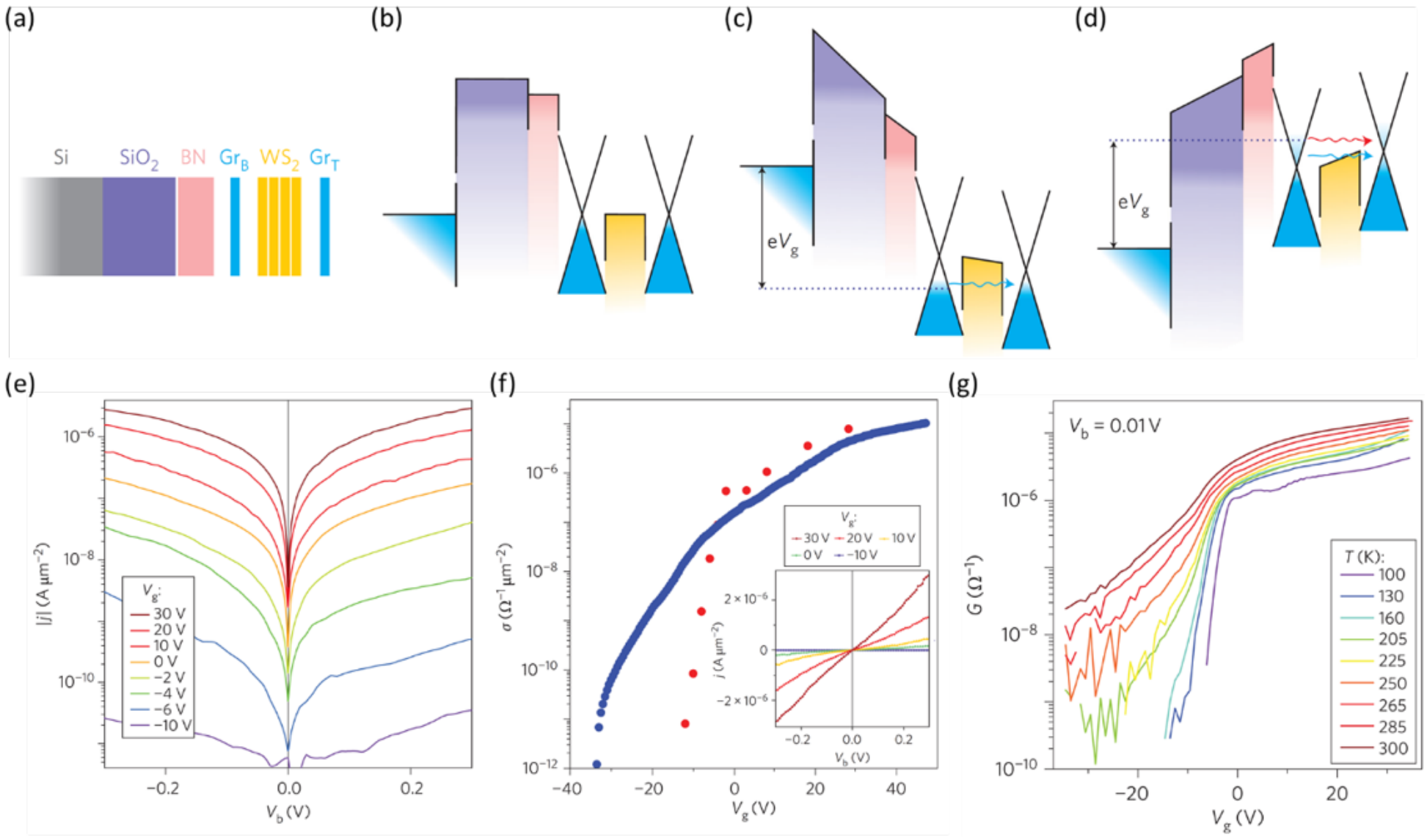

**Fig. 5** Tunneling FET based on Gr/TMD/Gr. (a) Schematic image of the device. (b)(c)(d) band alignments at zero gate voltage, negative gate voltage and positive gate voltage. (e) I-V curves at different back gate voltages at room temperature. (f) Gate dependent conductivity at room temperature. (g) Temperature dependent conductance. Reproduced with permission [80]. Copyright 2013, Springer Nature.

### 2.3 Gr/TMD/Metal vertical FETs

There is a trade-off between large ON current density and high ON/OFF ratio in the Gr/$MoS_2$/Gr and Gr/h-BN/Gr tunneling devices, due to the pinned Fermi level near top and bottom junctions. Replacing top graphene electrodes with metal (Fig. 6a) enables decoupling of gate tuning of top and bottom junctions[46]. Although the ON/OFF ratio drops from $10^6$ in Gr/TMD/Gr device to $10^3$ in Gr/TMD/metal device at room temperature (Fig. 6b), it is still suitable for logic application. In the meanwhile, the ON current debsity is increased from $2\times10^{-6}$ A/μm$^2$ to $5\times10^{-5}$ A/μm$^2$ (Fig. 6c), which mitigates the limitation of low ON current in the tunneling transistor. The increased ON current can be attributed to the thermionic emission or thermionic field emission processes, and thus the Gr/TMDs/metal vertical junctions operate like Schottky diode instead of tunneling barrier. As shown in Fig. 6d, when the bias is positive, the positive gate voltage leads to a decreased Schottky barrier height at Gr/$MoS_2$ interface and depletion width in TMDs layer, so that a large ON current is achieved by thermionic emission or thermionic field emission processes. Moreover, the ON current is almost constant due to the narrow depletion region, when the thickness of $MoS_2$ exceeds 10 nm (Fig. 6f). On the other hand, the use of negative gate voltage increases the Schottky barrier height and widens depletion region, resulting in a small OFF current. The thickness of TMDs thickness considerably affects the ON/OFF ratio. The ON/OFF ratio is increased as the TMDs thickness increases (Fig. 6g). It is worth to notice that the asymmetric I-V characteristic shows due toasymmetric gate tuning when the source and drain is changed. As shown in Fig. 6e, the electrons are injected into the junction through the metal/$MoS_2$ interface; however, the barrier height at metal/$MoS_2$ interface is unchanged with variation of gate voltage. Such asymmetric current is a typical result of Schottky-type barrier[142], which has been discussed in the $MoS_2$ transistor with two different Schottky contacts. As a general strategy of vertical transistor, we can always choose the sandwiched 2D materials to achieve different band alignments for specific functions. For example, the interlayer $MoS_2$ can be replaced by $Bi_2Sr_2Co_2O_8$ (BSCO) to make a transistor from n-type to p-type electrical transport, and realize vertical integration of a prototype complementary inverters based on metal/$MoS_2$/Gr/BSCO/Gr vertical vdW heterostructures. A semiconductor with smaller bandgap (*e.g.* BP) has been also used to form a vertical Gr/BP/Ni vdW FET [143]. Due to its small gap, a quite large ON-state current is reachable, however the ON/OFF ratio becomes much lower than that of Gr/$WS_2$/metal.

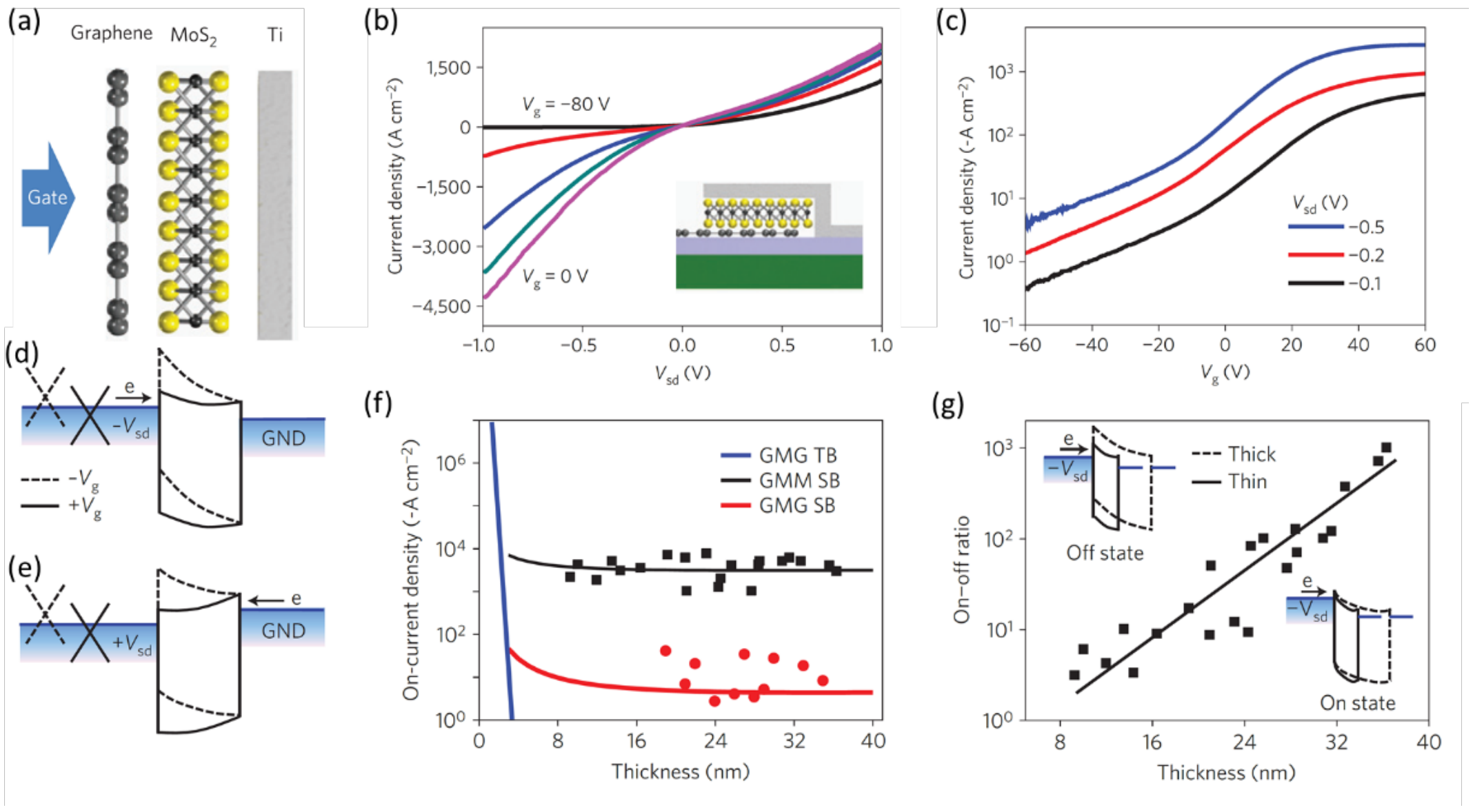

**Fig. 6** Vertical FET based on Gr/TMD/metal. (a) Schematic figure of device. (b) I-V curve with varied gate voltages. (c) Gate dependent current density with different bias voltages. (d) (e) band diagrams and mechanism of charge transport at positive and negative bias, respectively. (f) Thickness dependent ON and OFF current. (g) Thickness dependent ON/OFF ratio. Reproduced with permission [46]. Copyright 2013, Springer Nature.

In addition to thermionic emission mechanism, trap-assisted tunneling has been proposed in the Gr/TMDs/metal device structure to achieve large ON/OFF ratio and ON current density [144]. As shown in Fig. 7a, an ON/OFF ratio as high as $3 \times 10^4$ was reached at room temperature. As the temperature goes down to 160 K, the ON-state current will increase by a factor of ~ 4, while the OFF-state current will be reduced by more than 1000 times, leading to a largely enhanced ON/OFF ratio in low temperature (Fig. 7b). Different temperature dependence of ON-state current indicates the presence of two of more traps in $WSe_2$ (Fig. 7c). The defect states from W vacancies in the $WSe_2$ layers (11 nm thick) are supposed to determine the electronic structure of Gr/TMD interface (Fig. 7d) and assist the carrier tunneling process. Reduction in ON-state with increasing temperature indicates a trap-assist tunneling process. Although the ON/OFF ratio is much larger than that of the Gr/TMDs/Gr devices reported[46], the ON-state current is reduced by two orders of magnitude, indicating that the defect-assisted tunneling process is inefficient in terms of increasing ON current compared to thermionic emission process in the Schottky-type barrier,.

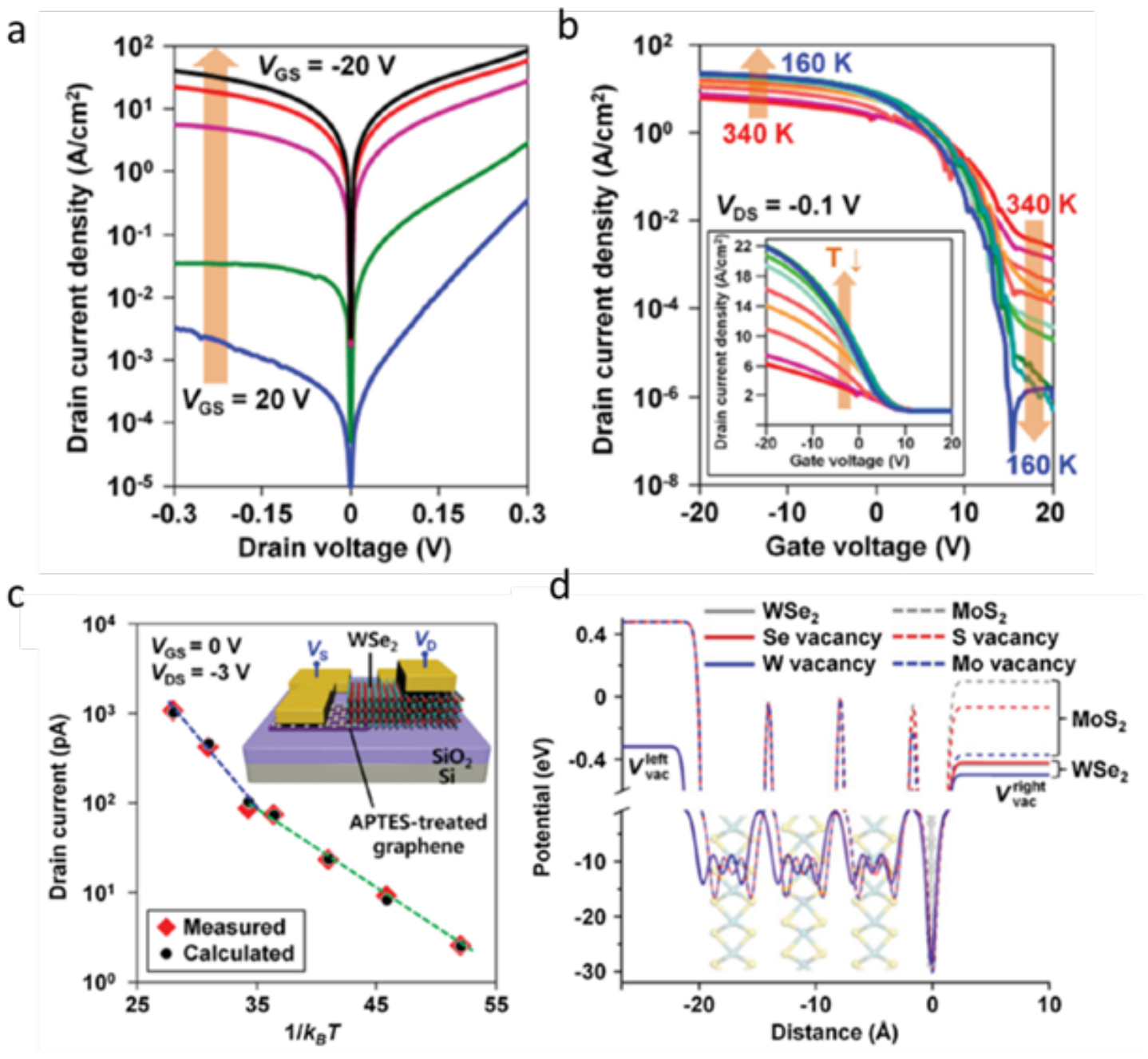

**Fig. 7** Trap-assist tunneling FET. (a) I-V curves for different gate voltages at room temperature. (b) Temperature dependent current density vs. gate voltage. (c) Temperature dependent current from measurement and theory. (d) Electrostatic potentials of the $WSe_2/MoS_2$ interfaces in different vacancy conditions from theoretical calculations. Reproduced with permission [144]. Copyright 2016, Wiley-VCH.

We have discussed several types of vertical transistors based on quantum tunneling. The achievable low OFF current is suitable for the low power applications. Moreover, the large ON/OFF ratio is usually attainable as long as the carrier transport mechanism is tuned from tunneling to thermionic emission for switching from OFF to ON states. However, the tunneling transistors have some shortcomings. First, the ON-current remains small for practical application, even other transport mechanisms like field emission from 2D materials[145, 146] have been demonstrated. Second, the performance strongly relies on the quality of insulating layer, and any defects or charge impurities may lead to current leakage in the OFF state. For example, the charge impurities in h-BN layer and local strains in TMDs layer will bring some "islands" in the insulating layer working as "quantum dot bridges" to assist the tunneling behavior, harming the performance of vertical tunneling transistors. Third, the electrically insulating 2D materials layers are too thick for wafer-scale integration. The h-BN with 4~7 layers is desirable to maximize the ON/OFF ratio while the semiconducting TMDs layers are required to be even thicker. However, so far there has been no developed techniques for growth of wafer-scale and high-quality thick h-BN or TMDs as the tunnel barrier. Fourth, the fabrication of vertical tunneling transistor requires the layer-stacking transfer technique, but the effect of the presence of bubbles or ripples cannot be excluded which will bring in local strains. Although in the lab-experiment level, clean area of sample after transfer can be chosen in the wafer scale fabrication process, the unavoidable defects or strain may limit the potential applications.

## 2.4 Vertical p-n junction TFETs

Vertical stacking of n-type 2D semiconductor and p-type 2D semiconductor allows for fabrication of vertical p-n junction TFET, which is also promising in low-power consumption. The most straightforward strategy is to stack two different monolayer semiconducting TMDs to form heterostructures with type II band alignment [93]. Figure 8a and b show the vertical p-n junction TFET based on n-$MoS_2$/p-$WSe_2$ and its current curves vs. bias voltage and gate voltage, respectively. The atomically thick vertical p-n junction devices (Fig. 8 c-d) show a gate-tunable rectifying behavior, which is due to the electrostatic modulation of carriers in the junction. Under a forward bias, tunneling-mediated interlayer recombination between electrons at the conduction band of $MoS_2$ and holes at the valence band of $WSe_2$ may be responsible for the carrier transport. When the recombination rate for Shockley-Read-Hall process and Langevin process is nearly balanced, the current is maximized without applying gate voltage. The current under forward bias is reduced with applying gate voltage.

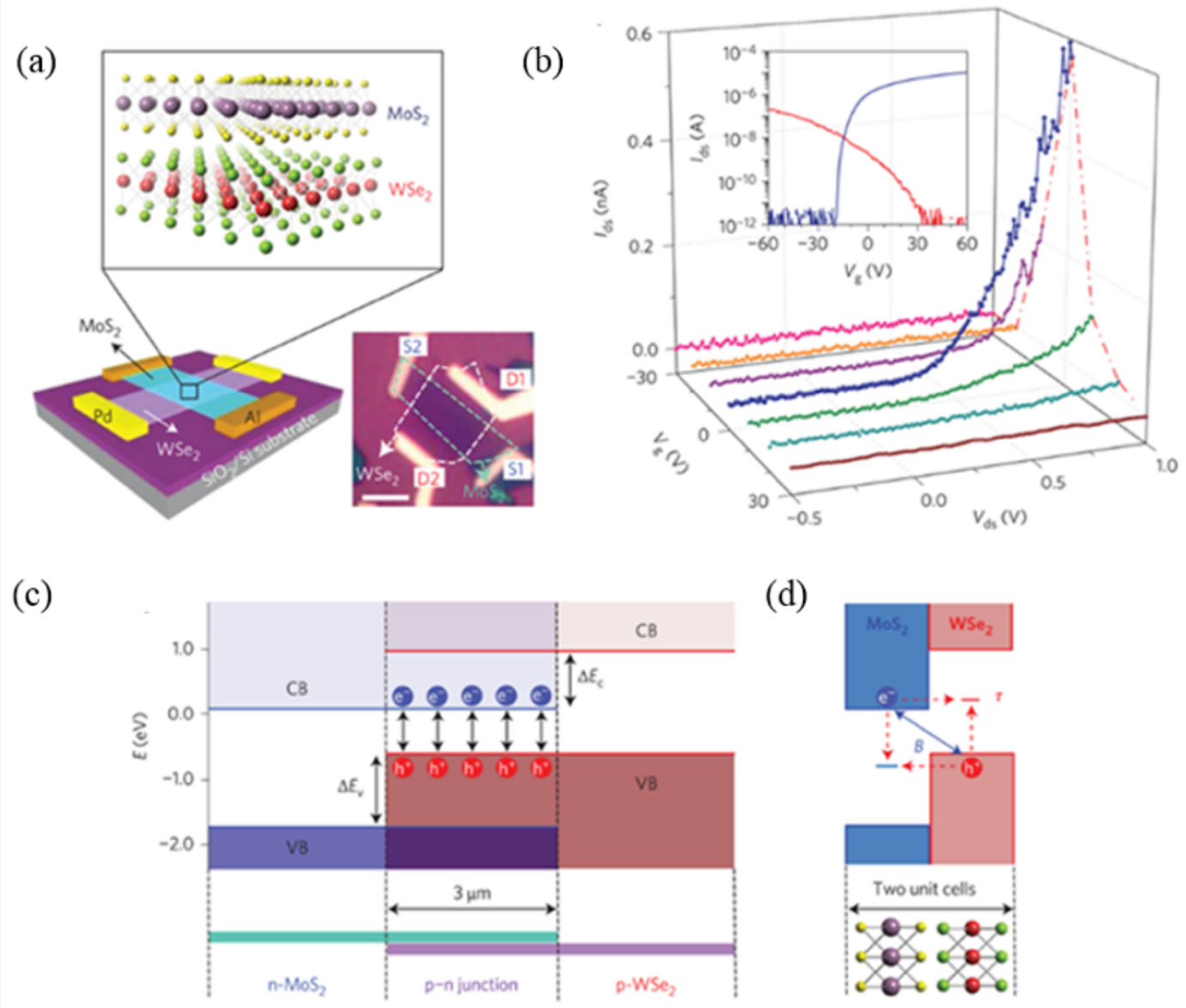

**Fig. 8** Charge transport across atomically thin p-n heterojunction. (a) schematic and optical image of $MoS_2$/$WSe_2$ heterostructure device. (b) Gate-tunable rectification of currentvoltages. Inset shows the transfer curves of $MoS_2$ and $WSe_2$ based FET devices. (c) band diagram for planar direction and (d) vertical direction in $MoS_2$/$WSe_2$ junction. Reproduced with permission [93]. Copyright 2014, Springer Nature.

Fueled by this pioneering work, different vertical vdW p-n junctions have been fabricated based on various 2D layered semiconductors. Apart from using TMDs as building blocks [147-150], the vertical p-n junctions based on other materials include BP/$MoS_2$ [85, 151], BP/$SnSe_2$ [152], GaTe/$MoS_2$ [153], $WSe_2$/$SnS_2$ [154], $WSe_2$/$SnSe_2$ [84, 155], BP/$ReS_2$ [156], $WSe_2$/GaSe[157], etc. According to the tunneling nature of the charge transport, the ON/OFF ratio in these p-n junction devices is relatively high ($10^3$~$10^7$). In addition, high rectification ratio can also be achieved in these junctions, but the origin of high rectification ratio is still controversial. Different from the semimetal graphene, semiconductors (*e.g.* TMDs and BP) are much more resistive than graphene at room temperature, thus semiconductor layer cannot be regarded as electrodes. This means that the tunneling current density across p-n junctions is not spatially uniform if metal electrodes are not deposited on top or bottom of the devices. In other words, a mixture of vertical and lateral tunneling usually takes place in the junctions. Considering the nature of tunneling across the p-n junction, the ON-state current is intrinsically low. But it is noted that the injected current is already small due to the large contact resistance at the source/drain terminals. Such intuitive understanding may be supported by the transport measurement of the metal/ $MoS_2$/BP/metal heterostuctures, in which the tunneling current is much larger than that of the devices with mixed in-plane and vertical transport[85]. Ideally, the subthreshold swing can be lower than 60 mV/dec in the TFET devices [158-160]. Recently a subthreshold swing of 55 mV/dec and 37 mV/dec has been realized in the BP/$MoS_2$ and $SnSe_2$/$WSe_2$ vertical p-n junction, respectively[82, 84]. To achieve much lower subthreshold swing, the tunnel junction bias and junction electric field should be optimized simultaneously[161]. However, this has been challenging since they are usually coupled. New or hybrid mechanism may be required to realize extremely small subthreshold swing in the vertical p-n junctions [162].

## 3. VdW heterostrucures for infrared photodetectors

Among diverse optoelectronic applications based on van der Waals heerostructures, infrared photodetectors have attracted considerable attention, as the vdW heterostructures offer unique properties such as high carrier mobilities and broad spectral absorption ranging from visible to infrared. [76, 163-165] High-performance infrared photodetectors based on vdW heterostructures are suitable for seamless monolithic integration with traditional Silicon technology due to free of dangling bond and excellent flexibility properties. When the infrared photodetectors are applied for practical imaging or communications, etc., sensitivity is one of key figures of merit, which can be manifested itself in external quantum efficiency, the responsivity and noise level of the infrared photodetectors. The external quantum efficiency accounts for the number of e-h pairs generated by the incident photon over the number of the incident photons per second: EQE $= I_{ph}/q\phi_{in}$, where $I_{ph}$ denotes the measured photocurrent, $q$ is the electron charge, $\phi_{in}$ is the incoming photon flux. The photoresponsivity can be determined by the ratio of photocurrent to the incident light power: $R_{ph} = I_{ph}/P_{in}$ , where $P_{in}$ corresponds to the light power. It is the combination of the responsivity and noise level that finally determines the sensitivity of the infrared photodetectors, in which the noise level is characterized by the noise equivalent power (NEP) in units of $W{\cdot}Hz^{-1/2}$. The specific detectivity characterizing the sensitivity can be expressed as $D^* = \sqrt{AB}R_{ph}/i_{noise}$ in the unit of Jones, where $A$ denotes the area of the photosensitive region, $B$ is the frequency bandwidth, $i_{noise}$ is the noise current at 1 Hz bandwidth. To make response to infrared light, bandgap of materials is required to be small. Recent efforts have identified few 2D materials with narrow bandgap, such as BP, b-AsP, $PdSe_2$ etc. These 2D materials including graphene are able to harvest infrared light to generate electron-hole pairs. By form vdW heterostructure with other 2D materials with wide bandgap, the light-generated carrier would be collected to achieve infrared light detection. Significant progress has been achieved along the improvement of room-temperature sensitivity of infrared photodetector with diverse vdW heterostructures. The specific detectivity of vdW heterostructure devices, *e.g.* about $10^9$ Jones for b-AsP/$MoS_2$ photodetector, can beat the commercial thermistor bolometer (about $10^8$ Jones) and PbSe MIR detectors (below $10^9$ Jones) at the mid-infrared wavelength.

**3.1 TMD-Gr-TMD vdW heterostructure**

For practical applications of photodetectors, it is highly desirable to have all key performance parameters competitive. The good photodetectors based on photoconductivity mechanism usually show ultrahigh photoresponsivity but extremely low response speed. Moreover, this type of photodetectors exhibits large dark current under a bias, leading to a low specific detectivity. Different from the photoconductive structures, the photovoltaic operating mechanism can be used to fabricate photodetectors without dark current and ultralow power consumption. High specific detectivity and ultrafast photoresponse speed are able to be achievable, due to rapid dissociation of excitons and suppressed noise in the photovoltaic junctions. By inserting 1L or 2L graphene into the p-n junction made of $MoS_2$ and $WSe_2$ as shown in Fig. 9a, Long et al has demonstrated a p-g-n based infrared photodetector with ultrafast photoresponse and high specific detectivity [76]. In this structure, the infrared light is absorbed by graphene to generate e-h pairs. The vertical channel length of the p-g-n device is comparable to the vdW gap between $MoS_2$ and $WSe_2$ and therefore it is very short. Strong built-in electric field can be created in the junction and the photogenerated e-h pairs in the graphene will be rapidly separated by the strong field (Fig. 9b). As a result, noise from generation-recombination process is reduced and the p-g-n photodetector shows a high specific detectivity in the near-infrared region (Fig. 9c). Due to the weak optical absorption of the graphene, the external quantum efficiency is low in the p-g-n structure. To increase the external quantum efficiency, one solution is to increase optical absorption by increasing the thickness of graphene. By inserting thicker graphene into p-n junction[166], a h-BN/$MoTe_2$/7L-graphene/$SnS_2$/ h-BN infrared photodetector has been fabricated through a layer-by-layer transfer technique (Fig. 9d and 9e). This type of p-g-n device is highly sensitive at room temperature (Fig. 9f). Compared to the performance the $MoS_2$/Gr/$WSe_2$ device, the photoresponsivity and specific detectivity of h-BN/$MoTe_2$/7L-Gr/$SnS_2$/ h-BN photodetector were much improved (Fig. 9g). However, the response speed becomes slower than $MoS_2$/Gr/$WSe_2$ structure due to the use of thicker

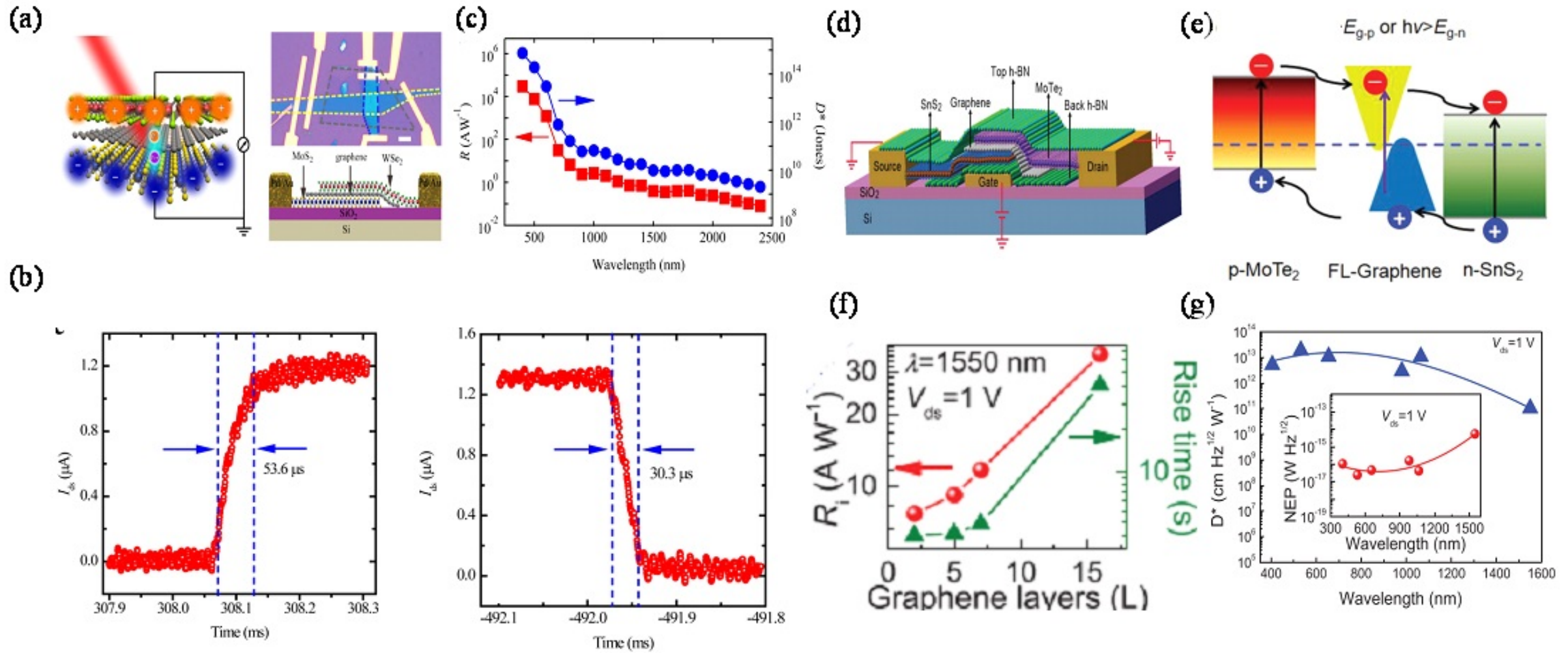

**Fig. 9** p-g-n infrared photodetection and performance (a) schematic of $MoS_2$/Gr/$WSe_2$ device structure (b) response speed (c) photoresonsivity and specific detectivity of $MoS_2$/Gr/$WSe_2$ infrared photodetector versus wavelength (d) schematic of h-BN/$MoTe_2$/7L-Gr/$SnS_2$/h-BN device structure (e) its band diagram (f) photoresponsivity and rise time as a thickness of graphene (g) specific detectivity and NEP (inset) as a function of wavelength. (a-c) Reproduced with permission [76]. Copyright 2016, American Chemistry Society. (d-g) Reproduced with permission[166]. Copyright 2016, Wiley-VCH.

graphene in the junction region (Fig. 8f). Therefore, increasing the thickness of graphene is infeasible to improve the overall performance of infrared photodetector based on p-g-n structure.

Thin black phosphorus (BP) film has very narrow band gap and may be used as a better optical absorber to increase the external quantum efficiency and achieve higher specific detectivity [167]. Although the performance of infrared photodetectors based on p-g-n is not as good as expected, these results have indicated huge potential of vertical vdW junctions in realization of infrared photodetector with high-sensitivity and fast response speed.

### 3.2 BP or B-AsP/TMD vdW heterostructure

The bulk BP is a semiconducting material with narrow bandgap (about 0.3 eV) [21, 168-170], which is increased to about 2 eV in its single layer form[171]. Use of bulk BP allows for the detection of the mid-infrared wavelength up to 4.13 um. To detect the long-wavelength infrared, semiconductors with much narrower bandgap are desirable. By alloying arsenic element with black phosphorus, the synthesized b-AsP shows the good tunability of bandgap[172, 173], with results demonstrated in Fig. 10a. The bandgap is reduced with increasing the arsenic chemical content. When the ratio of As to P is 0.83:0.17, the band gap is reduced down to 0.15 eV. Based on this materials, a van der Waals photodetector has been built by stacking p-type b-AsP semiconductor on top of n-type $MoS_2$ to detect the mid-infrared light [174]. The formation of p-n junction is justified by the observation of rectifying behaviors in Fig. 10b and the photocurrent mapping without the bias. The photodetector shows good photo-responsivity and EQE (Fig. 10c), both of which are dependent on the wavelength of detection. The photogenerated electron-hole pairs at the p-n junction under the light illumination can be separated rapidly to contribute to the measured photocurrent. The resulting photoresponse of this kind of photodetector is very fast (Fig. 10d). The rising/falling time is within the scale of microseconds. Besides, the presence of potential barrier at the junction suppresses the noise resulted from the generation-recombination and thermal effects, leading to considerable performance improvement over commercially available room-temperature infrared detectors, *e.g.* PbSe-based photodetector and thermistor (Fig. 10e). Compared to the ex-situ doping (i.e. alloying), in-situ tuning of the band gap is more suitable for detecting infrared light over the broad spectrum. The band gap can be tuned over a wide range by the vertically electric field. A h-BN/BP/h-BN heterostructure mid-IR photodetector with dual gate configuration has been demonstrated[175]. This photodetector operates under the photoconductive mechanism. Applying a vertical electric field not only alters the oscillator strength to shift light absorption edge at the charge neutrality regime (Fig. 10f), but also shrinks the bandgap to increase free intrinsic carrier density to suppress the photoresponse, which show a widely tunable spectrum response up to 7.7 μm. The maximum responsivity for 5 nm-thick BP film is estimated to reach 28.7 A/W at 7.7 μm after considering the optical absorption of Pt top gate and BP. The NEP at 3.4 μm operating at charge neutrality point is able to reach 0.03 $pWHz^{-1/2}$, which shows better performance than the photodetectors based on b-AsP phototransistor and b-AsP/$MoS_2$ heterojunction. The photoresponsivity and NEP are expected to be further improved by integrating the thin BP with optical structures e.g. cavity and waveguide. Due to the short lifetime of photogenerated carriers less than nanosecond, the operation speed of BP-based photodetectors goes beyond 1 GHz [175]. By further reducing the lifetime, larger bandwidth may be achieved and is beneficial to the high-speed optical communications. The optical absorption of BP and b-AsP takes place through inter-band transitions. This leads to a modest responsivity and a challenge in detecting weak optical signals. To solve this issue, these 2D material can be coupled to optical nanostructures to enhance the light-matter interactions. The narrow bandgap of semiconductor enables the photodection of mid-IR wavelength, but it also poses a key challenge in generating dark current through thermal generation of carrier, which is significant for room-temperature mid-IR photodetectors. Optimizing the thickness of sensing materials may offer a potential avenue for reducing the dark current to improve the performance of photodetectors. For BP- or b-AsP-based mid-IR photodetectors, the optimal detectivity has been achieved within the thickness range of 25-35 nm-thick (Fig. 10g), where increasing optical absorption and reducing dark current are balanced. [177]

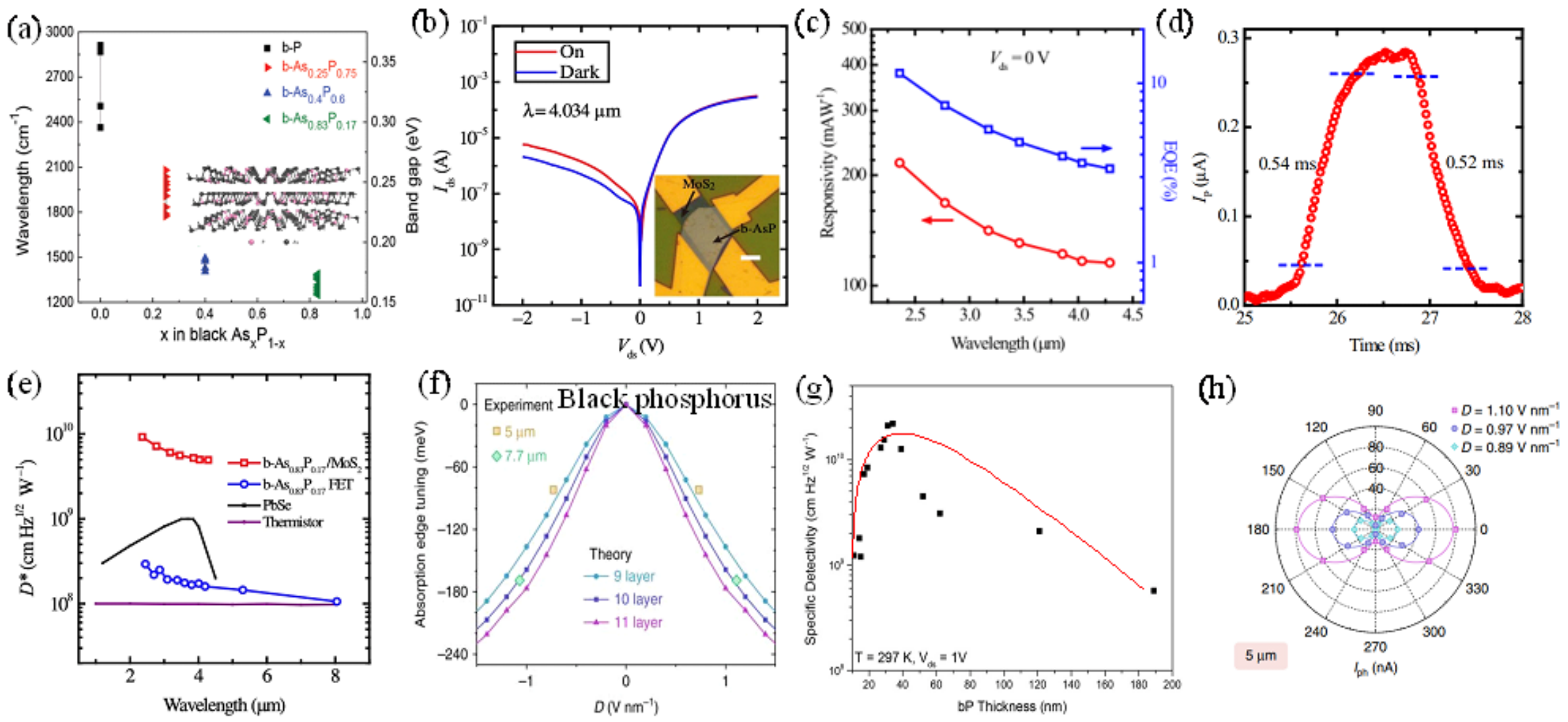

**Fig. 10** Infrared photodetection based on black phosphorus and black arsenide phosphorus. (a) the band gap and wavelength versus composition variation of arsenide (b) Rectifying characterization of B-AsP/$MoS_2$ heterostructure photodetector with and without mid-infrared light illumination (c) responsivity and external quantum efficiency as a function of wavelength (d) rising and falling time (e) comparison of specific detectivity of infrared photodetector based on different materials (f) variation of absorption edge with perpendicular electrical displacement (g) thickness dependence of specific detectivity (h) anisotropic properties of black phosphorus based infrared photodetector with change of electrical displacement. (a) Reproduced with permission [172]. Copyright 2016, Wiley-VCH. (b-e) Reproduced with permission [176]. Copyright 2017, American Association for the Advancement of Science (AAAS). (f,h) Reproduced with permission [175]. Copyright 2017, Springer Nature. (g) Reproduced with permission[177]. Copyright 2017, American Chemistry Society.

The polarized light scattered from an object usually includes rich information of the object, such as surface roughness and geometry [164], etc. Resolving the polarization contrast is helpful in identifying the target through contrast enhancement in optically scattered environments. Due to the difference in optical absorption along the different crystallographic orientations, BP and b-AsP are capable of resolving light polarization[178], which enhances the functionalities of mid-IR detectors and expands their applications. Most interestingly, the responsivity of photodetector based on BP or b-AsP strongly depends on the polarization of infrared light. Applying an electrical field along out-of-plane of BP or b-AsP allows for sensitive detection of polarization (Fig. 10h). The sensitive polarization, broad spectral response and high photodetectivity as well as fast response speed make black phosphorus and b-AsP promising candidate materials used for mid-IR photodetectors.[179]

Previous works have shown the potential of the BP- and TMDs-based heterostructures in mid-IR photodetection. However, the external quantum efficiency is less than 5% and still inferior to the commericialized room-temperature photodetector[179]. Recently, Bullock et al have fabricated the photodiode based on BP/$MoS_2$ heterostructure and achieved the external quantum efficiency of exceeding 30% over the wavelengths ranging from 2.5 to 3.5 $\mu$m. With the high quantum efficiency, a high peak detectivity ($1.1 \times 10^{10}$ cm $Hz^{-1/2}$/W) at wavelength of 3.8 $\mu$m without bias has been obtained. By using the anisotropic optical property in the puckered lattice, they demonstrated a new type of the polarization-resolved mid-IR photodetector that enables simultaneous detection of orthogonally polarized light without using external polarizer."

**3.3 TMD/TMD vdW heterostructure**

Photodetectors based on TMD/TMD vdW heterostructures have been demonstrated, however, the large bandgap for conventional TMDs restricts their photoresponse to visible light spectrum. Recently, the group 10 family of noble metal dichalcogenide has been discovered for tunable bandgap, high mobility and air stability[180-186]. These important properties can bridge the gap between the graphene and black phosphorus to enable the detection of mid-wavelength infrared. Among the promising candidates are $PtSe_2$ and $PdSe_2$. Both have a band gap of around 1.2 eV in its monolayer and become semimetal in the bulk form. The widely-tunable band gap with the thickness endows the photodetector based on these materials with the capability in achieving broad spectrum photodetection. By fabricating $PtS_2/PtSe_2$ heterostructure devices, photodetection of broad band spectrum ranging from 405 nm to 2200 nm without applying the voltage bias [187] was demonstrated. The photovoltaic mechanism allows for the operation of self-driven infrared photodetector, indicating the possibility of devising photodetectors with extremely low-power consumption. The use of ultrathin $PtSe_2$ in the infrared photodetectors leads to large dark current and low specific detectivity due to the low optical absorption. Using the thin 2D $PtSe_2$ film to form heterojunction with CdTe, which is an II-VI group direct-semiconductor with a bandgap of around 1.5 eV, a near-infrared photodetector was demonstrated to show a fast response speed of microsecond scale at room temperature [188]. Compared to the $PtSe_2$-based heterostructure infrared photodetectors, the photodetector based on $PdSe_2$ heterostructure shows better performance in terms of sensitivity and specific detectivity.

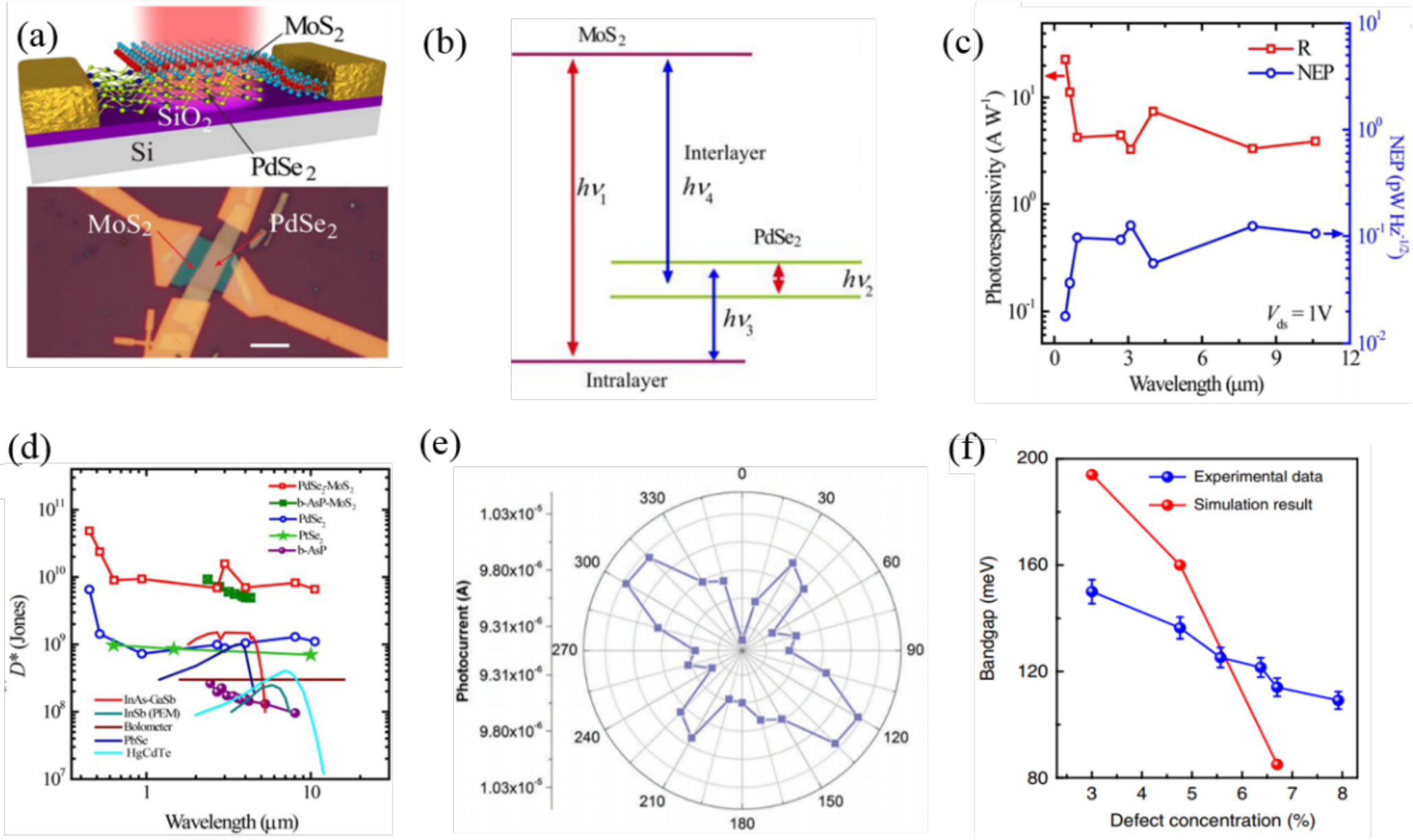

**Fig. 11** (a) $PdSe_2/MoS_2$ infrared photodetector and optical image of corresponding device (b) schematic illustration of exciton absorption in $PdSe_2/MoS_2$ heterostructure (c) photoresponsivity and NEP as a function of wavelength (d) comparison of specific detectivity (e) polarization dependence of photocurrent in $PdSe_2$ device (f) band gap of $PtSe_2$ versus defect concentration. (a-d) Reproduced with permission[189]. Copyright 2019, American Chemistry

Society. (e) Reproduced with permission [190]. Copyright 2019, Wiley-VCH. (f) Reproduced with permission [184]. Copyright 2018, Springer Nature.

Stacking n-type $MoS_2$ onto of p-type $PdSe_2$ enables the fabrication of a p-n junction infrared photodetector [189] (Fig. 11a). This photodetector shows broad spectrum photoresponse ranging from visible light (450 nm) to long-wavelength infrared (10.6 μm). Over the mid- to long-wavelength infrared spectrum, the photoresponsivity of the photodetector is not changed too much. Interestingly, the maximum photoresponsivity occurs at 4 μm. The high optical absorption at 4 μm may be attributed to form of interlayer exciton (Fig. 11b). The exact mechanism calls for more work to clarify. Similar to the other p-n junctions, the noise can be efficiently suppressed. The $PdSe_2/MoS_2$ photodetector shows a NEP of lower than 0.13 $pW/Hz^{-1/2}$ over the broad infrared spectrum (Fig. 11c). With the low noise level, the specific detectivity at 4 μm reaches $6.09\times10^{10}$ Jones at room temperature. This value is comparable to that of b-AsP/$MoS_2$ photodetector and well goes beyond the infrared photodetectors based on graphene and $PtSe_2$. This performance even beats the uncooled HgCdTe and commercial bolometers (Fig. 11d). Unlike the BP- or b-AsP-based photodetectors, the performance of $PdSe_2/MoS_2$ photodetector is not degraded after one-year exposure to dry air. With the unique pentagonal lattice structure, $PdSe_2$ shows anisotropic optical absorption of incident light similar to BP or b-AsP, paving the way for serving as a linear dichroism media (Fig. 11e). In addition to tuning band gap by varying thickness, the band gap of these noble metal dichalcogenides can be modified via the defects engineering to access the photodetection of far-infrared spectrum (Fig. 11f), which decreases with increasing defect concentration [184]. Together with high performance, air stability and polarization sensitive, the photodetector based on $PdSe_2$ vdW heterostructure may offer a platform for more practical applications in remote sensing and thermal/medical imaging.

### 3.4 BP/InSe vdW heterostructure

Sensitivity and noise level are two key parameters for photodetection of single photon, where the high sensitivity and low-noise level are required. The photodetectors based on photoconductivity or photovoltaic mechanisms have been facing challenge in achieving the high sensitivity and low-noise level. Instead, the photodetectors based on avalanche phenomenon can be used for this purpose. The avalanche photodetector is based on semiconductor p-n junction and operates under a reverse bias. The carrier multiplication induced by the acceleration of photogenerated carriers in the electric field is responsible for the operating mechanism. For the avalanche photodetector based on traditional materials, e.g. Si, Ge, a very large reverse voltage (exceeding tens of volts) is applied across a long active channel region to achieve carrier multiplication with high photogain, which limits their practical applications to specific areas.

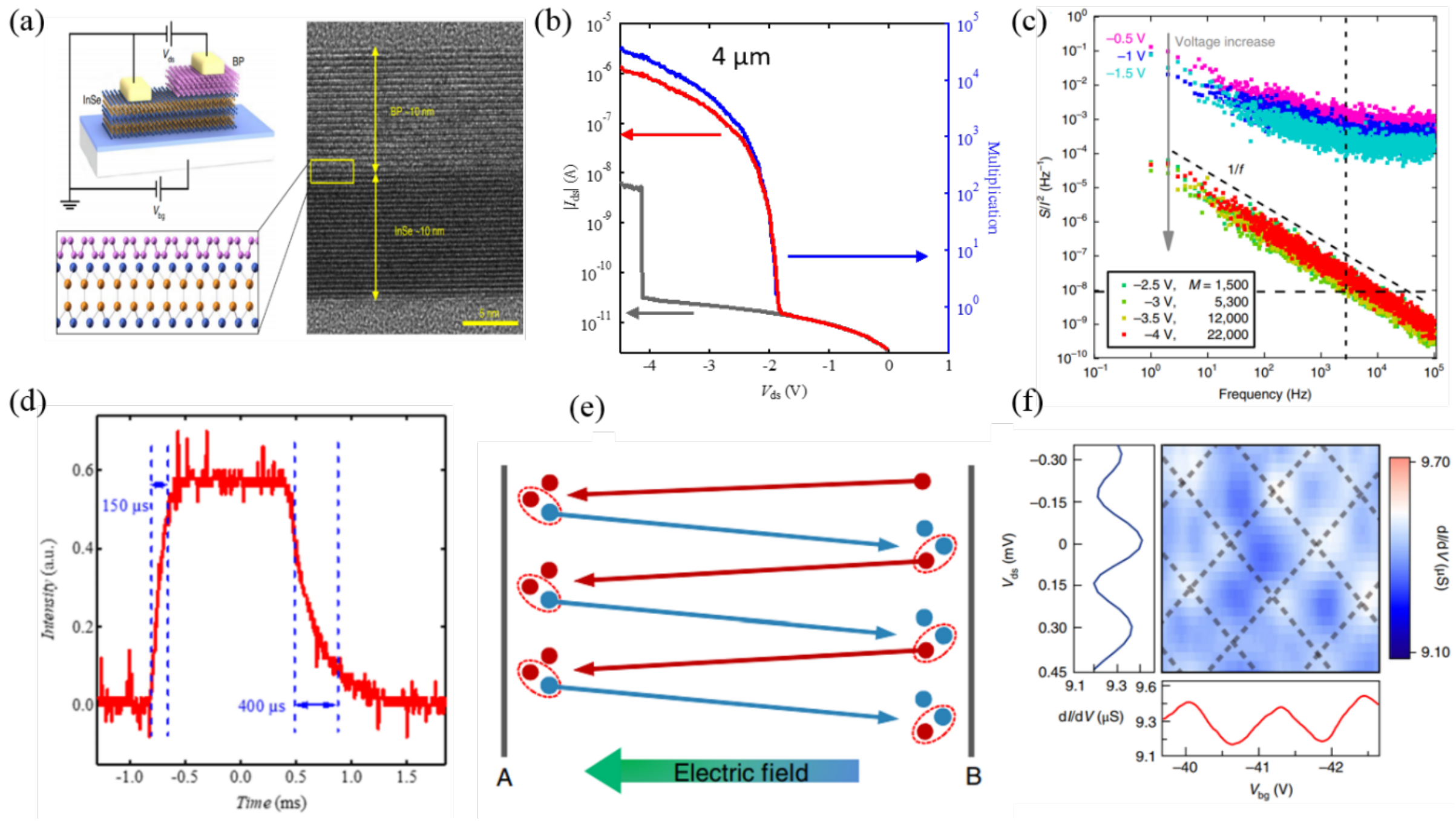

**Fig. 12** (a) Schematic of BP/InSe vertical heterostructure based mid-infrared avalanche photodetector and the cross-section of high-resolution tunneling electron microscopy of BP (b) I-V curves with and without mid-infrared light (c) noise spectrum for different bias voltages (d) response speed (e) schematic of ballistic transport and impact ionization of electrons and holes (f) the evidence of ballistic transport of carrier along the out-of-plane of BP. Reproduced with permission [162]. Copyright 2019, Springer Nature.

Vertical vdW heterostructures made of p- and n-type two-dimensional materials have high-quality and very sharp interface and are promising in realization of high-performance avalanche mid-infrared photodetector. By stacking p-type BP and n-type InSe exfoliated through mechanical transfer approach into vertical heterostructure[162], Gao *et al.* has fabricated an avalanche photodetector with highly sensitive at mid-infrared detection of 4 μm (Fig. 12a). With the whole assembling processes in a glove box, the vdW interface is very clean as corroborated by the cross-section of high-resolution transmission electron microscopy. Under the action of reverse voltage, current suddenly increases by 5 orders of magnitude. The infrared illumination on the vdW heterostructure device reduces the avalanche voltage down to 1 eV. The photon-gain reaches $10^3$ (Fig. 12b) and is much larger than the avalanche diodes based on the conventional materials at the same bias. This photodetector is very powerful and is capable of detecting the ultra-weak light of about 50 pW, corresponding to 6000 incident photons. The capability of this type of photodetector is not limited to this level. Theoretically, its detection limit may push to 0.38 pA, lower than that of commercial photodetectors based on HgCdTe. The effect of avalanche on the current noise density is profound. The noise level is drastically reduced in the regime of avalanche and the noise shape is different from the excess noise of traditional avalanche photodetectors (Fig. 12c). Interestingly, the amplified carrier density does not contribute to the current fluctuation, indicating its huge potential in the detection of single photon. In addition to the ultrahigh sensitivity, the photoresponse of the avalanche photodetector based on the vdW heterostructure is very fast and the time scale is within the scale of microseconds (Fig. 12d). Such superior performance of this photodetector is attributed to the ballistic avalanche transport occurring in the BP and its high-quality interface with InSe (Fig. 12e-f). A hole accelerated in the electric field without any scattering to generate an e-h pair at the interface of BP and InSe. Two holes can be collected, and the generated electron will be reversely accelerated to generate another e-h pair at the interface of BP and metal electrode. Afterwards, more and more carriers will be generated and then collected at the interface, which account for the ballistic avalanche process.

The infrared photodetection is promising in imaging systems and memory devices. Individual photodetectors based on van der Waals heterostructures have demonstrated excellent performance and some key performance parameters are even superior to the commercial products as reviewed above. However, it has been challenging to achieve the integration of large device array based on the individual device. Photodetectors with high photo-gain show huge potential in the high-quality imaging applications. The development of imaging systems has evolved from the line arrays to focal plane arrays and from cooled to uncooled types. With further advance in infrared technology, the concept for the fourth generation of infrared photodetector has been proposed[191], which emphasizes the functionalities of higher resolution, polarization and multicolor. After extensive research efforts, the room-temperature performance of individual devices based on 2D materials has been improved to be comparable with the traditional thin films. It is critical to assess the feasibility of assembling the photodetectors based on 2D vdW heterostructures into large arrays for practical applications and their performance. By integrating the graphene-quantum-dots based photodetectors with CMOS electronic read-out circuits, a $388 \times 288$ large photodetector array has been demonstrated [192]. The successful demonstration of high-resolution imaging based on the monolithic integration of vdW heterostructure photodetector opens up a promising path towards vertical integration of two-dimensional materials with distinct electronic and optoelectronic functions in the circuits. In addition to imaging, the infrared photodetection shows promising application in the neuromorphic computing, since light as a stimulus can be captured to induce resistance change of devices. The radiation of matter in general covers the whole infrared spectra, detection of which has been widely involved in the night vision, medical diagnosis and identification of military targets. Vertical heterostructure integrates few 2D materials with distinct electronic and optical properties and shows versatile band alignments, which may realize infrared memory devices [193].

## 4. VdW heterotructure for spintronic applications

The vertical vdW heterostructures are not limited to the electronic and optoelectronic device applications. The advent of 2D vdW magnetic crystals makes the diverse vdW heterostructures promising in the spintronic applications, including memory storage devices and spin TFETs. One of most typical applications related to spin is the hard disk drives (Fig. 13a) based on magnetic tunneling junctions (MTJ)[194]. The MTJ is a vertical magnets/insulator/magnets heterostructure device. The tunneling resistance is different when the two magnets have parallel moments or anti-parallel moments. Such difference in tunneling resistance is characterized by tunneling magnetoresistance (TMR) ratio, which is defined as the tunneling resistance of parallel magnets over that of anti-parallel magnets. The giant MTR ratio arises from the asymmetric dependence of spin polarization in the two magnetic layers. Spin is polarized in the magnetic electrodes of the MTJ and the density of states (DOS) for spin up and spin down at Fermi level are different (Fig. 13b) [194]. Such spin polarization results in the asymmetric tunneling resistance for parallel and anti-parallel configuration of spins in the two magnetic electrodes. Over the past decades, the room-temperature TMR ratio has been continuously increasing [195], and is expected to reach 800% at 2027 and 1000% at 2032 (Fig. 13c). Note that conventional MTJ devices suffer from the non-uniform tunneling across the tunnel barrier. Such issue can be avoided by replacing the conventional tunel barrier materials with 2D layered materials (*e.g.* h-BN, graphene) with atomic flat and clean interface. While 2D magnetic materials have been pursued for years in the hope of realizing low-power consumption and compact spintronic applications, 2D magnetism described by Heisenberg model won't give rise to finite Curie temperature. In order to get non-zero Curie temperature, magnetic anisotropy is necessary to break the rotational symmetry in Heisenberg model.

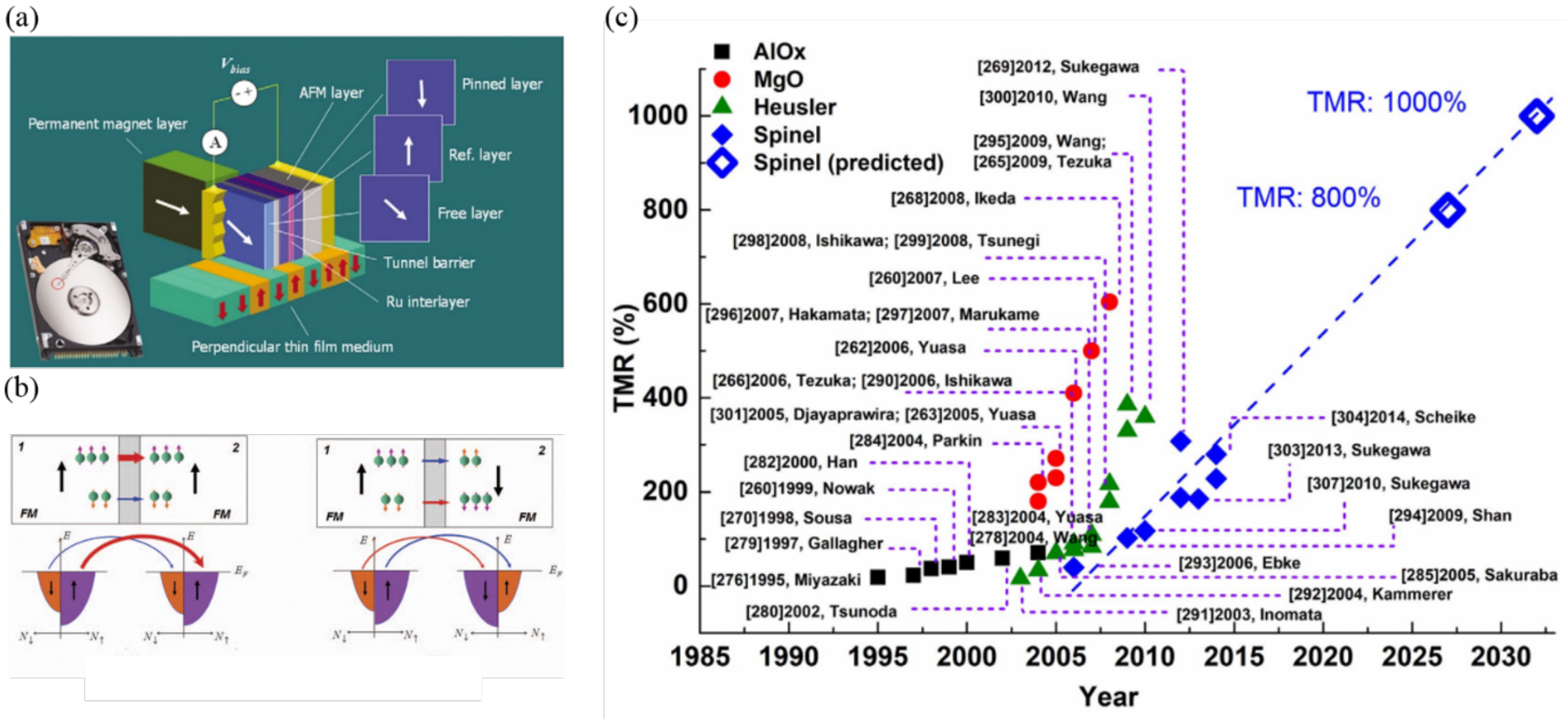

**Fig. 13** (a) Schematic image of of MTJ operations in Hard Disk Drive. (b) The mechanism of TMR in the MTJ. (a,b) Reproduced with permission [194]. Copyright 2006, Elsevier. (c) The increasing curve of TMR vs. years, reproduced with permission [195]. Copyright 2019, IEEE.

Discovery of intrinsic magnetism in vdW materials down to strict 2D limits has further fueled the dreams of low-power and ultra-compact spintronic device applications. These 2D magnetic materials include insulating $Cr_2Ge_2Te_6$ (CGT) [28] and $CrI_3$ [29], metallic $Fe_2GeTe_2$ (FGT) [25] and $VSe_2$ [196]. As shown in Fig. 14a [28], CGT can be mechanically exfoliated into few layers. The magnetism at different temperatures has been characterized by the Magneto-optic Kerr effect (MOKE) under a finite magnetic field of 0.075 T, which introduces a small magnetism anisotropy. A small magnetic field can introduce an anisotropy term into the Heisenberg model to enhance the 2D magnetism and increase the Curie temperature. Thickness dependence of Curie temperature (Fig. 14b) of few-layer CGT indicates that the interlayer coupling is able to enhance the magnetism. Moreover, the magnetic hysteresis loops is gate-tunable (Fig. 14c), which has been claimed to be due to the moment rebalance in the spin-polarized electronic band structure of few-layer CGT [26]. Similar to $Cr_2Ge_2Te_6$, 2D $CrI_3$ exhibits ferromagnetic order below the temperature of 45 K. However, interlayers are coupled with antiferromagnetic order [29], which is different from $Cr_2Ge_2Te_6$. Figure 14e shows thickness-dependent magnetic hysteresis loops [29]. The intralayer exchange interaction in monolayer $CrI_3$ gives rise to a ferromagnetic order. The antiferromagnetic order coupling in bilayer $CrI_3$ leads to three steps in magnetic hysteresis loop. In trilayer sample, the ferromagnetic order recovers due to a non-vanishing net magnetism. The controllable spin alignment among different layers opens up a novel avenue for devising the memory devices base on vdW magnetic tunneling junctions. 2D vdW magnets with high Curie temperature is

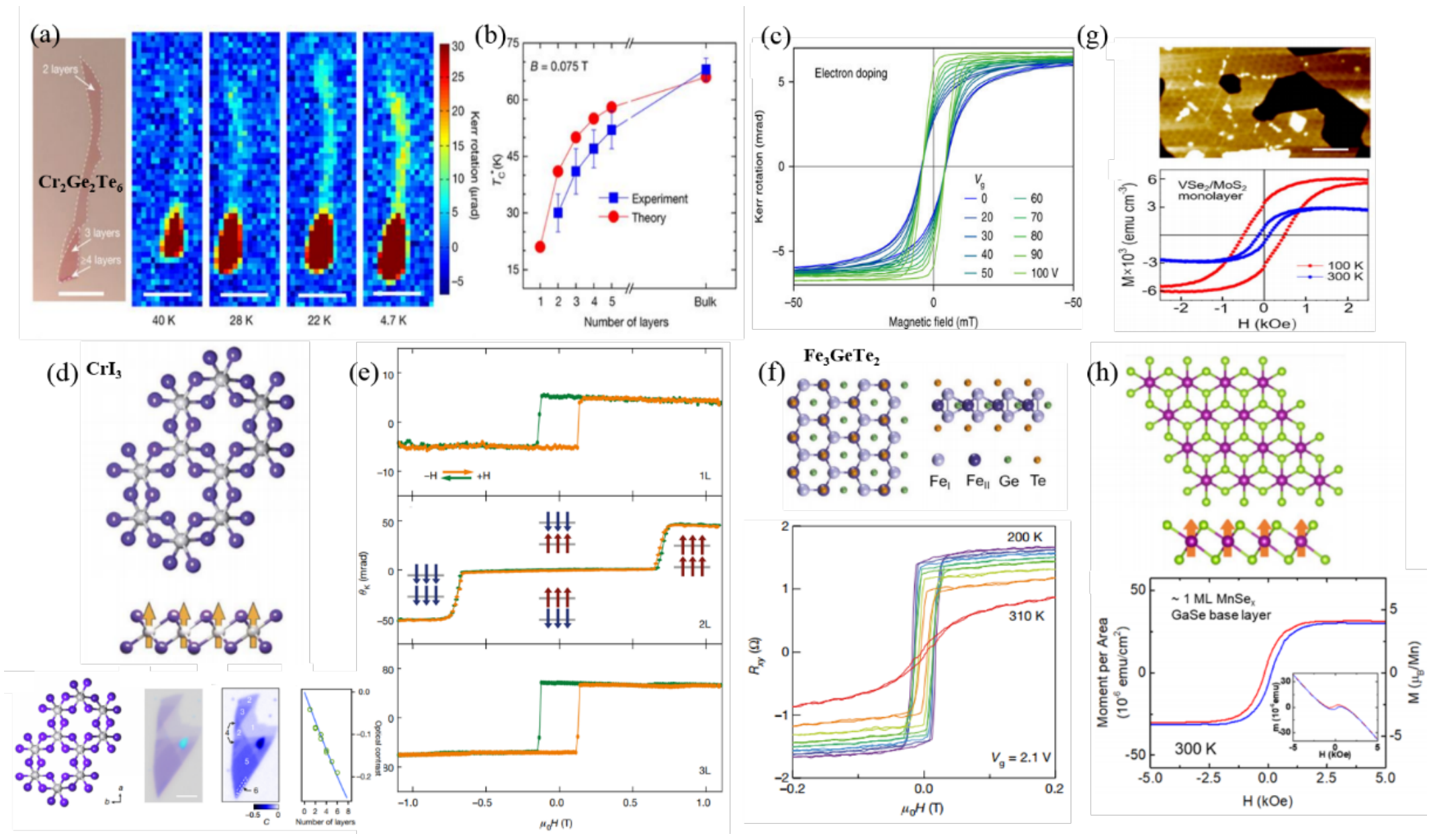

**Fig. 14** (a) optical image and Kerr rotation intensity mapping in the few layer $Cr_2Ge_2Te_6$. (b) Curie temperature vs. number of layers for theory and experiment, respectively. (a, b) Reproduced with permission [28]. Copyright 2017, Springer Nature. (c) Magnetic hysteresis loop at electron doping in $Cr_2Ge_2Te_6$, reproduced with permission [26]. Copyright 2018, Springer Nature. (d) Schematic image of crystal lattice and magnetization in monolayer $CrI_3$ as well as Optical image and contrast map of few layer $CrI_3$. (e) Magnetic hysteresis loop at mono-, bi- and tri-layer $CrI_3$. (d, e) Reproduced with permission [29]. Copyright 2017, Springer Nature. (f) Schematic image of crystal lattice of $Fe_3GeTe_2$, and the temperature-dependent magnetic hysteresis loop, reproduced with permission [25]. Copyright 2018, Springer Nature. (g) STM measurements of $VeSe_2$ monolayer and magnetization hysteresis loop at 100K and 300K, reproduced with permission [196]. Copyright 2018, Springer Nature. (h) Schematic structure and magnetization image of $MnSe_x$. Magnetic hysteresis loop of monolayer $MnSe_x$ at room temperature, reproduced with permission[197]. Copyright 2018, American Chemical Society (ACS).

desirable for practical spintronic applications. However the Curie temperature of 2D magnetic insulators is far below room temperature and obstructs the promising application of 2D vdW magnets. Encouragingly, Curie temperatures close to or even above room temperature become achievable in several 2D metallic magnets. Mechanically exfoliated $Fe_3GeTe_2$ represents a typical example, whose crystal structure is illustrated in Fig. 14f [25]. The anomalous Hall signal can be used to characterize the magnetism. At the temperature of 310K, the survival of a weak anomalous Hall signal indicates a room-temperature 2D ferromagnetism. Besides, molecular beam epitaxy grown 2D materials, such as $VSe_2$ on HOPG[196] (Fig. 14g) and $MnSe_x$ on GaSe [197] (Fig. 14h), exhibit room-temperature ferromagnetism,. However, the origin of ferromagnetic order in these materials remains open and calls for more effort to clarify.

**4.1 Gr/h-BN heterostructure based spin devices**

Since graphene has negligible intrinsic spin–orbit coupling[198] and small hyperfine interactions as well as extremely-high carrier mobility[3], it is very promising in transporting spin information over a long distance and has the potential to be used for future development of new types of spintronic devices[199]. Aside from the long spin diffusion length, efficient spin injection and detection also play key role in realization of high-performance spintronic applications like spin transistor[200]. For Gr/ferromagnetic metal (FM) based spin valve devices, impedance-mismatch issue usually leads to very low efficiency of spin injection through Gr/FM contact[201-203]. To solve this problem, a tunnel barrier is required to insert into between graphene and ferromagnetic metal electrode to suppress the back flow of spin current from graphene[204]. The quality of the tunnel barrier and the interface with graphene considerably affect the efficiency for spin injection. To be specific, the presence of pinholes in the tunnel barrier would increase spin absorption and accelerate spin relaxation. The nonuniform interface caused by deposition of barrier materials enhances the spin-flip scattering. Choosing different barrier materials and deposition techniques partially alleviates these issues, the details for which have been reviewed in the ref. 206 [205]. Fully solving the issues associated with spin injection calls for effort in seeking tunnel barrier materials with free of pinholes and atomical flatness as well as highly resistive resistance. 2D layered-structure h-BN material is the desirable one, as it has a wide bandgap of about 6 eV and exhibits electrically insulating property and strong dielectric breakdown (around 1 V/nm) in its monolayer form. Theoretically, it has been proposed that 100% spin injection efficiency can be achievable in the Ni/thick h-BN/Gr heterostructure[206]. Experimentally, only 1-2 % spin polarization[207] in the bilayer graphene was reported with single-layer h-BN as tunnel barrier, which may be due to the low resistance at out-plane direction of monolayer h-BN. Subsequent efforts have increased the spin polarization up to 65 % by use of 2-3 layers h-BN[208]. This can be attributed to high resistance and suppressed proximity effect of polymer residues on the graphene interface. The spin polarization injected through Gr/h-BN /FM may be further improved by fabricating high-quality interface. Although much promising results on spin injection and transport in the graphene have been achieved, further efforts in understanding of spin relexation mechanism in the graphene and overcoming the technology challenges (*e.g.* gate-tunabl spin-orbit coupling) are required to make the practical Gr/h-BN heterostructure based spintronic device applications possible[209].

**4.3 FGT/h-BN or graphite/FGT MTJs**

Van der Waals magnetic tunneling junctions based on pure 2D materials hold promise in overcoming the non-uniform tunneling issue in the conventional MTJ devices, offering an ideal platform to realize high magnetic resistance ratio. Based on newly discovered 2D vdW magnets, several types of vdW magnetic tunneling junctions (MTJ) have been demonstrated. With h-BN sandwiched by two $Fe_3GeTe_2$ electrodes with different shapes, $Fe_3GeTe_2$/h-BN/$Fe_3GeTe_2$ junction [107] acts as a spin-valve device (Fig. 15a), with the spin polarization configurations corresponding to the high resistance state and low resistance state shown in Fig. 15b. The TMR ratio can reach 160% at 4.2 K. Different alignments of magnetic moment in the top and bottom electrodes generate distinct I-V curves. I-V curve is linear for low resistance state, as the barrier height and width of the thin layer h-BN won't change as the bias is swept. While, it becomes non-linear for high resistance state (Fig. 15c). In this case, the TMR ratio shows a peak at zero bias and is drastically reduced with deviation from zero bias. The bias dependent TMR ratio is a ubiquitous behavior in the MTJs, but the origin is still not clear.

Graphene is another tunneling material used in the vdW devices [210]. Note that a non-magnetic metal was used to separate the two magnetic metals in early GMR device. Different from conventional metals, few-layer graphene has atomically-flat and dangling-bond-free surface. These properties would help suppress the spin-flip scattering when the spin tunnels through the barrier. The optical and AFM image of FGT/graphite/FGT device is shown in Fig. 15d. Compared to the FGT/h-BN/FGT device, the TMR ratio is much smaller in FGT/graphite/FGT device. The most unusual phenomenon is the asymmetric TMR with sweeping magnetic field, which is opposite from the normal MTs. Such asymmetry in tunneling resistance is independent of the thickness of the graphite layer. Just like FGT/h-BN/FGT device, the positions of jumps in the field dependent TMR ratio indicates the flip of the spin polarization in the magnetic electrodes. In order to determine the spin polarization configuration, the selected thicknesses for the top and bottom FGT electrodes are different, so that FGT layer exhibiting spin flipping can be recognized according to the amplitude of jump. Together with the angle dependent measurement (Fig. 15e), the origin of asymmetric TMR ratio was identified. When two FGT layers have antiparallel spin polarization, two possible configurations exist: (i) upwards spin polarization in the top layer FGT, while downwards spin polarization in the bottom layer FGT, which is defined as "OUT" state; (ii) downwards spin polarization in the top layer FGT, while upwards spin polarization in the bottom layer FGT, which is defined as "IN" state. "IN" and "OUT" states have opposite effect on the TMR ratio of FGT/graphite/FGT device. This is different from that of normal MTJ devices. Most surprisingly, such effect on TMR ratio depends not only on the "IN" or "OUT" configuration, but also the current direction.

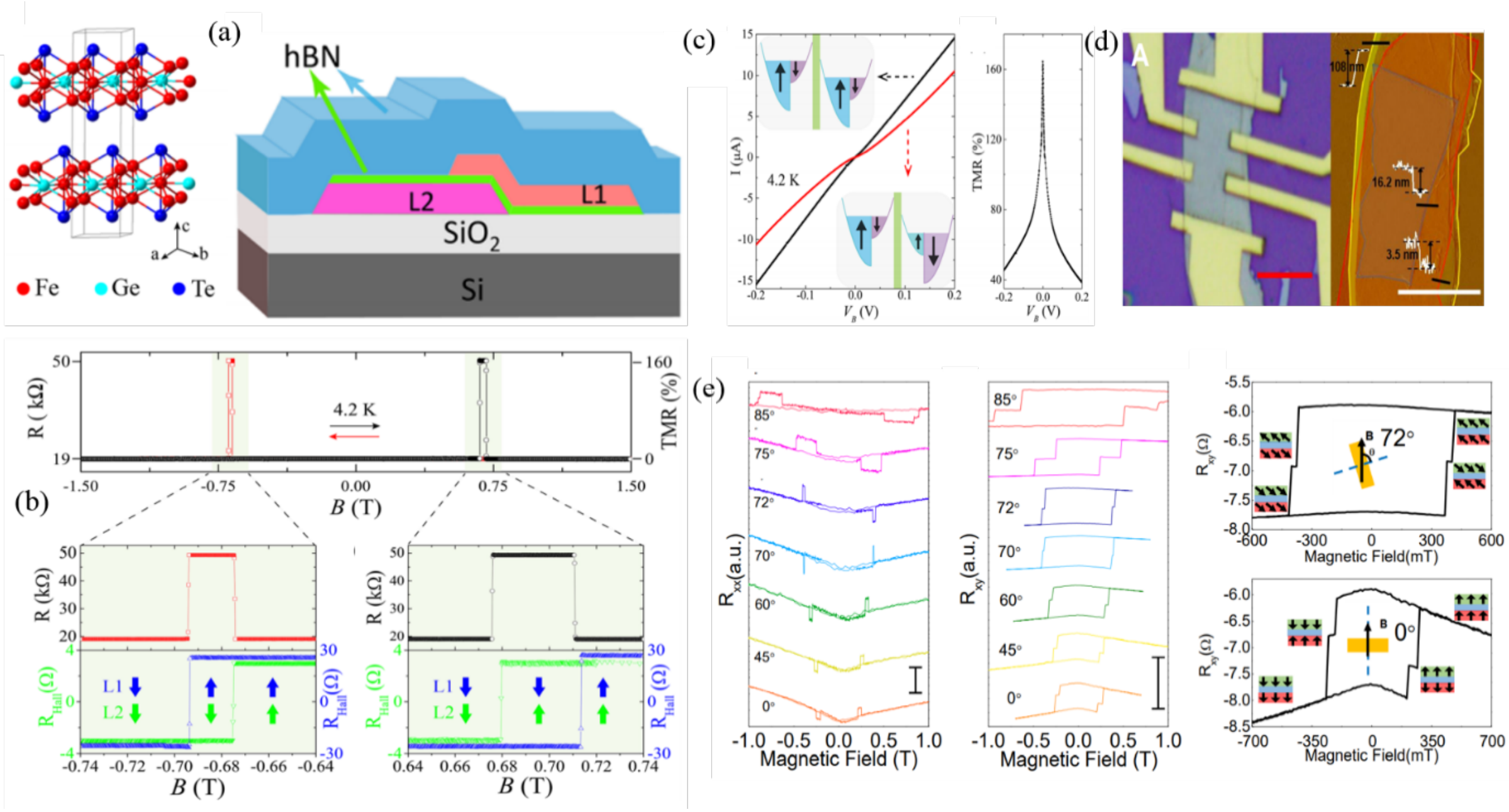

**Fig. 15** vdW MTJ device based on FGT. (a) Schematic image of FGT structure and FGT/h-BN/FGT device. The magnetoresistance is measured at 4.2K. (b) Spin valve effect in FGT/h-BN/FGT device. (c) Bias dependence of magnetization configuration and TMR ratio. (a, b, c) Reproduced with permission [107]. Copyright 2018, American Chemical Society (ACS). (d) Optical and optical image of FGT/graphite/FGT device. (e) Angle dependent $R_{xx}$ and $R_{xy}$ at 20K. The correlations between the angle, Rxy and magnetization configures are illustrated. (d, e) Reproduced with permission[211].

### 4.4 Gr/$CrI_3$/Gr MTJs

Stacking of 2D vdW magnet and 2D non-magnetic conductors allows for devising novel MTJ devices, in which 2D vdW magnetic insulator is sandwiched by two non-magnetic 2D conducting electrodes. This type of devices may be promising in low-power storage application and highly-sensitive control of the magnetic tunneling junctions. One typical Gr/$CrI_3$/Gr device has vertical geometry with its optical image shown in Fig. 16a, in which the $CrI_3$ are selected to be monolayer or bilayer [110]. With varying electrostatic doping of $CrI_3$, monolayer and bilayer $CrI_3$ show distinct magnetism transition. The main difference for monolayer and bilayer $CrI_3$ is that monolayer $CrI_3$ is interlayer ferromagnetic while bilayer $CrI_3$ has an interlayer anti-ferromagnetic spin polarization, which has been verified by the magnetic circular dichroism (MCD) measurement at 633 nm using a confocal microscope. When the gate voltage is applied, the saturation magnetization, coercive force and Curie temperature of monolayer $CrI_3$ tunneling device are largely tuned. On the other hand, in the bilayer $CrI_3$ tunneling device, the magnetic field responsible for transitions between FM and AFM phase shows a shift with varying carrier concentration. Particularly, the hysteresis loop with gate sweeping displays different field effect behavior for different magnetism strength of bilayer $CrI_3$. The mechanism for gate-tunable magnetism of mono- and bilayer $CrI_3$ is still unclear, although several theories have been proposed [212, 213].

In addition to electrostatic doping, the vertical electric field can also cause the magnetic transition in bilayer $CrI_3$ [111]. Unlike monolayer $CrI_3$, bilayer $CrI_3$ has two spatially distinct magnetic layers in the vertical direction, and as a result, the vertical electric field can push the carriers to transfer between these two layers. The electric field dependent MCD measurements show linear magnetoelectric (ME) response from 4 K to about 60 K, and the linear slopes of ME curves are respectively positive, zero and negative for positive ferromagnetism, anti-ferromagnetism and negative ferromagnetism under positive electric field. The positive and negative directions are defined by two opposite directions perpendicular to the $CrI_3$ plane. As shown in Fig. 16b, vertical electric field induces the transition between ferromagnetic state and anti-ferromagnetic state through ME coupling. Such a strong ME coupling in Gr/bilayer $CrI_3$/Gr device suggests that the device may be used for electrically controlled nonvolatile memory, spintronic, etc.

Several different configurations of net magnetic orders can be achieved dependent on the spin polarization for each $CrI_3$. Replacing the magnetic tunneling layer with trilayer or four-layer $CrI_3$ (Fig. 16c) layer leads to the presence of additional steps in the magnetic hysteresis loop. For one net magnetism state in devices with trilayer or four-layer $CrI_3$, the presence of a few distinct spin polarization configurations (Fig. 16d) would lead to much larger TMR ratio [73] than TMJ devices based on monolayer or bilayer $CrI_3$. For example, the TMR ratio largely increases to 3200% in trilayer $CrI_3$ TMJ device and future increases to 19,000% in four-layer $CrI_3$ TMJ device. This is due to the spin-filtering effect through interlayers with anti-ferromagnetic coupling order. Similarly, TMJ device based on tri- or four-layer $CrI_3$ exhibits gate-tunable current and TMR ratio (Fig. 16e). Most unusual thing is that even the net magnetism is unchanged, the current or TMR ratio can still be changed due to change of the spin polarization configuration[113], (Fig. 16f), pointing to a potential application in magnetically moderated transistor with low power.

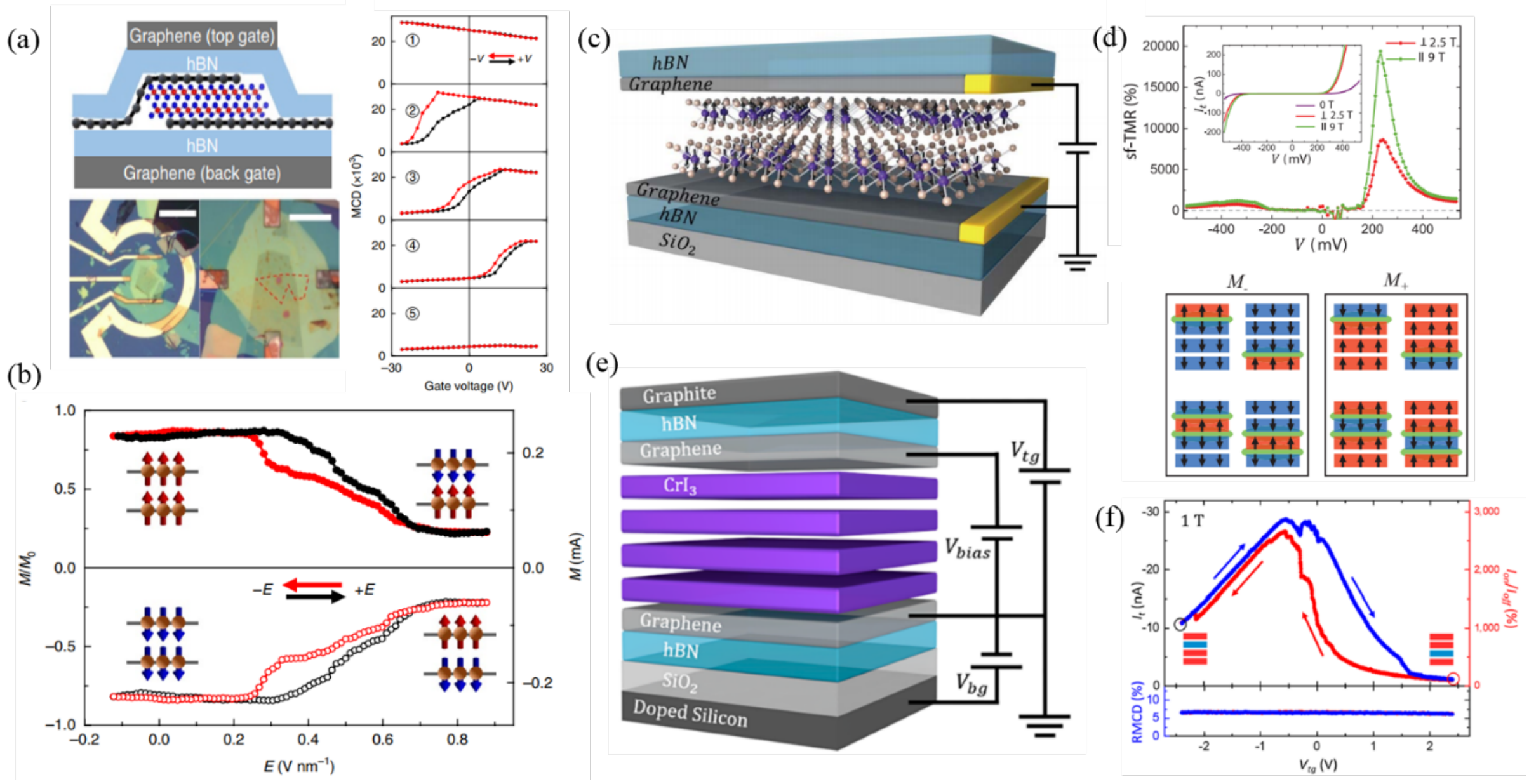

**Fig. 16** vdW TMJ device based on $CrI_3$. (a) Schematic image of vertical Gr/$CrI_3$/Gr TMJ with top gate and back gate, and the magnetization of $CrI_3$ tuned by electric doping. (b) Magnetization of $CrI_3$ tuned by vertical electric field. (a, b) Reproduced with permission [110, 111]. Copyright 2018, Springer Nature. (c) Schematic image of spin filter MTJ device based on $CrI_3$. (d) the TMR ratio vs. bias voltage and the transition of magnetization configuration in spin filter MTJ device. (c, d) Reproduced with permission [73]. Copyright 2018, The American Association for the Advancement of Science (AAAS). (e) Schematic figure of another example of vdW spin filter MTJ device. (f) The TMR ratio is tuned by the gate in spin filter MTJ device. (e, f) Reproduced with permission[214]. Copyright 2019, American Chemical Society (ACS).

### 4.5 Spin Tunnel Field-Effect Transistors

As discussed above, the MTJ devices based on vdW heterostructures give rise to a high TMR ratio compared to conventional MTJs devices. The property of gate-tunable TMR ratio makes vertical vdW MTJ devices more promising. In addition to MTJ devices, another promising application of vdW MTJ is the spin tunneling field effect transistor. In the traditional spin field-effect transistor, the spin transport is controlled by gate-tunable spin-orbit coupling, while spin field effect in the 2D vdW magnetic MTJ devices is based on different mechanism. The magnitude of tunneling current is controlled by the gate-tunable transition between different magnetic orders. For example, for spin tunnel field-effect transistors based on bi-layer Gr/$CrI_3$/ bi-layer Gr TMJ structures (Fig. 17a)[108], FM and AFM spin polarization configurations in thicker $CrI_3$ layer is switchable by gate voltage, rendering $CrI_3$ layer serving as a spin filter for tunneling. The operating mechanism is distinct for the spin tunneling field-effect transistor based on bi-layer Gr/$CrI_3$/ bi-layer Gr. Applying gate voltage changes the spin polarization configuration of multilayer $CrI_3$ and affects transitions between high resistance state and low resistance state (Fig. 17b). Therefore, the operating mechanism relies on gate-controlled spin filtering effect instead of band alignment, indicating the promise of low power operation. With such distinct working principle, high ON/OFF current ratio is attainable in spin tunneling field-effect transistors, pointing to new proposal for possible non-volatile and reconfigurable memory and logic applications.

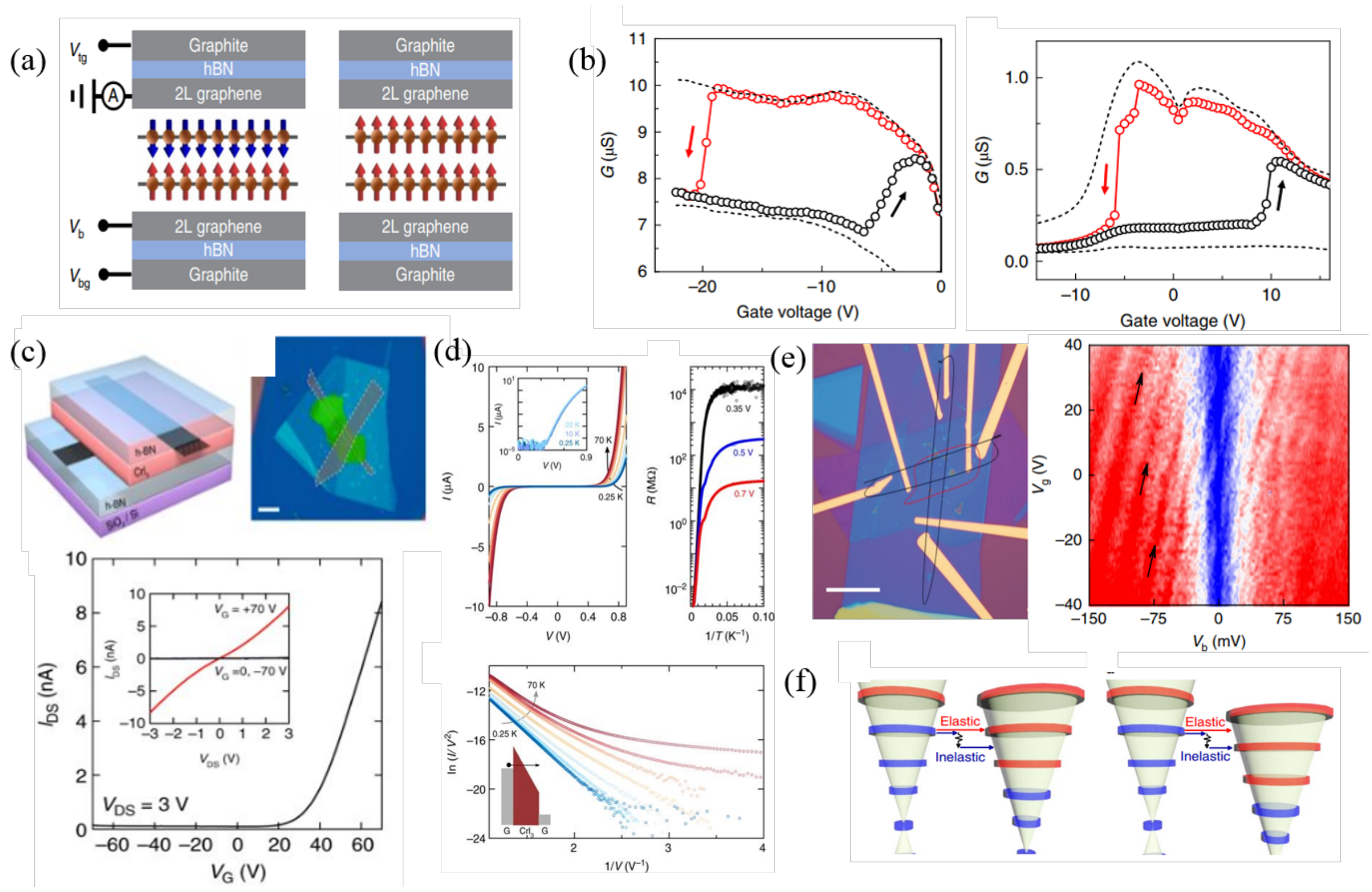

**Fig. 17**. Spin tunneling FET device. (a) the schematic image with the anti-ferromagnetic and ferromagnetic configuration. (b) The conductance loop with gate sweeping. (a, b) Reproduced with permission[108]. Copyright 2019, Springer Nature. (c) Schematic and optical image of spin tunneling FET based on $CrI_3$. The field effect transport of only $CrI_3$ is shown as well. (d) the tunneling results and mechanism of the device shown in (c). (c, d) Reproduced with permission [114]. Copyright 2018, Springer Nature. (e) Optical image of Spin tunneling FET device based on $CrBr_3$. The gate and bias voltage dependence of transport properties at finite magnetic is shown. (f) the mechanism of magnon-assist tunneling. (e, f) Reproduced with permission [215]. Copyright 2018, Springer Nature.

Similar magnetic order dependent tunneling field effect transistor behavior has been also reported [114]. Planar field-effect transistor with $CrI_3$ as the channel (Fig. 17c) shows semiconductor-like field effect characteristics with high ON/OFF ratio. In the vertical junction, the different spin polarization configurations have different impact on the tunneling current due to the spin filtering effect. Apart from the spin filtering effect, the effect of band alignment tunneling current is also profound. The reason is that the $CrI_3$ layer itself works as a bias-dependent tunneling barrier to assist Fowler-Nordheim tunneling behavior, which can be justified by temperature dependent transport behaviors (Fig. 17d). In the vertical TFETs, phonon or charge impurity facilities tunneling through resonant scattering. Similarly, it is possible to transfer momentum in the spin tunneling process in MTJs [215]. Recently, magnon- and magnetic-impurity-assisted spin tunneling has been reported in the Gr/$CrBr_3$/Gr vertical heterostructures (Fig. 17e-f), which was justified by the magnon-assisted inter-Landau level tunneling under the external magnetic field.

## 5. Challenges and Outlook

For electronic device applications, vdW sandwich structure and p-n junction heterostructures have been widely explored as a promising platform to implement electronics applications in the tunneling field effect transistors and gate-tunable Schottky barrier vertical transistors as well as band-to-band tunneling transistors. Some key challenges are limiting the further technology advance of achieving superior-performance electronics devices on bar with or even outperforming the state-of-art silicon-based field effect transistors. The first challenge is lack of complete device physics governing the carrier transport across the vertical heterostructure or vertical p-n junction. For example, Richardson's law developed for bulk materials is usually utilized in the prior experimental works to account for the thermionic emission mechanism of carrier transport over the Schottky barrier of graphene and 2D semiconductors. Note that the thickness of 2D atomically-thin crystals is comparable or even smaller than de Broglie wavelength. It is therefore inappropriate to apply this bulk transport law to vdW heterostructures. Although some revised theories or models [216-220] incorporting the effect of band structures of 2D materials have been proposed to understand the unique carrier transport across atomically-sharp vdW heterostructures, it has been difficult to distinguish the correct physical picture only by fitting the experimental data with the new models or theories. Besides, the new semiconductor physics models used to describe the carrier transport across the vertical p-n junction are required to develop as the thickness of 2D materials is much smaller than the depletion width. Due to the lack of understanding the carrier transport across the vdW heterostructures, it is difficult to perform full simulation for vertical heterostructure based devices by using multi-scale device simulators like T-CAD to achieve the optimized device parameters, which is very helpful to fabricate high-performance electronic devices.

For infrared photodetection devices, it is important to achieve the fast speed and highly sensitive infrared photodetection in order to realize imaging applications. The atomically-thin 2D materials used as channel of photodetectors offer high mobility and gate-tunable Fermi level and are promising in realization of high photoresponse speed. The large dark current poses a big challenge in achieving high specific detectivity, although the presence of potential barrier appearing in the interface of 2D vdW heterostructures can suppress the noise current level to some extent. Currently, the record-high specific detectivity of the infrared photodetectors is achieved at the cost of very slow response speed at milliseconds or even seconds, which severely limits the application where the fast speed and high sensitivity are desirable. The challenge faced by the infrared photodetector is how to achieve the high sensitivity without sacrificing other key figures of merit.

For the spintronic device applications, vdW heterostructures have shown promising device concepts, although 2D intrinsic magnetic crystals were just discovered by either mechanical exfoliation or molecular beam epitaxial approach. Two challenges have to be overcome before practical technological advance. For example, the operation of spin-filter MTJ devices based on magnetic vdW heterostructures requires low temperature and the use of magnetic fields. However, the room-temperature MTJ devices with nonvolatility and low-power consumption are desirable in the practical applications. To solve this challenge, much effort has to be devoted to seeking 2D vdW magnets with high Curie temperature and perpendicular anisotropy. Besides, it is also challenging to achieve electrical control of reversal between distinct magnetic orders and lower critical current for spin-orbit torque magneto resistive random access memory among vdW heterostructure-based spintronic applications.

Large-area growth of vdW heterostructures is also a challenge that limits device applications[43]. Currently, the fabrication of vdW heterostructures mainly relies on mechanical transfer technique. Mechanical transfer of 2D materials was first reported in 2010 for Gr/h-BN heterostructures[221]. Such transfer process is only suitable for fabricating micron-scale sample. Although other methods have been developed subsequently to realize the transfer for large-scale CVD samples[222-224], it is still a challenge to achieve direct synthesis of vdW heterostructures with distinct materials on $SiO_2$ substrate[17, 225-227]. Besides, developing devices with diverse functionalities is important at post-Moore era. Nevertheless, fabrication of multi-functional vdW heterostructure devices is not easy since it involves the complex mechanical transfer and vertical stacking of distinct 2D materials[39, 228-231]. So far, the demonstration of multi-funcational devices mainly relies on the complicated structures. The multiple mechanical transferring processes involved in the fabrication of devices definitely affect the performance of devices and may lead to low device yields. Moreover, most of the multifunctionalities demonstrated are based on charge transport and photon, implying that the power consumption may be still a concern of large-scale integration of multifunctional devices.

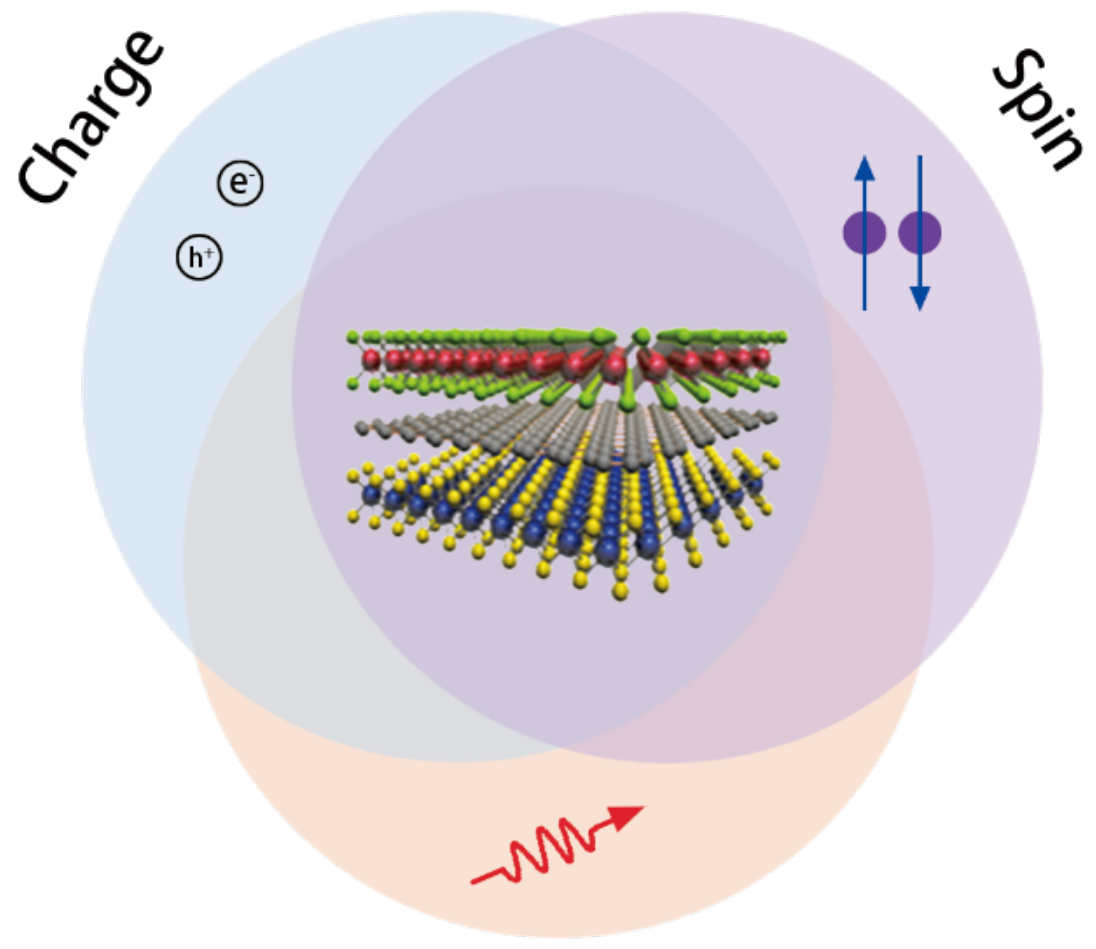

**Fig. 18** Vision of multi-functional vdW devices through coupling of distinct degrees of freedom.

The challenges also provide us with opportunities. With relentless search for new 2D materials, the family members of 2D materials has been continuously growing and more material selections are available in stacking vdW heterostructures. The library of vdW heterostructure materials will be further expanded, with the degree of freedom in stacking consequence [17, 46, 93, 222, 224, 225, 232, 233]. As the fabrication technology of vdW heterostructure becomes advanced, it would be possible to fabricate much complex vdW heterostructure materials. Based on these sophisticated structures, it is likely to design novel multifunctional devices utilizing the degrees of freedom (*i.e.* charge, spin and photon) inaccessible in the existing material systems as shown in Fig. 18, which is one of very promising research direction in the future. For example, reconfigurable solar spin devices may be achievable by stacking p-type vdW ferromagnetic semiconducting material with n-type nonmagnetic 2D semiconducting material[234]. It is also conceivable to employ spin hall effect and spin orbit torque of vdW heterostructure, which is made of 2D ferromagnetic metal and 2D metallic material with strong spin orbit coupling, to design a device capable of processing in memory[235-237]. Considering practical applications, integrating the advantages of vdW devices with mature 3D material devices based on the standard fabrication processing may offer a promising path for future technology advance.

## Acknowledgements

This work was supported in part by the National Key Basic Research Program of China (2015CB921600), the National Natural Science Foundation of China (61974176, 61625402, 61574076), and the Collaborative Innovation Center of Advanced Microstructures and Natural Science Foundation of Jiangsu Province (BK20180330 and BK20150055), Fundamental Research Funds for the Central Universities (020414380122, 020414380084).

Received: ((will be filled in by the editorial staff))
Revised: ((will be filled in by the editorial staff))
Published online: ((will be filled in by the editorial staff))